\documentclass{aa}  
\usepackage{graphicx}
\usepackage[]{txfonts}
\usepackage{hyperref}

\begin{document}

   \title{Abundances of disk and bulge giants from \\high-resolution optical spectra\thanks{Based on observations made with the Nordic Optical Telescope (programs 51-018 and 53-002), operated by the Nordic Optical Telescope Scientific Association at the Observatorio del Roque de los Muchachos, La Palma, Spain, of the Instituto de Astrofisica de Canarias, spectral data retrieved from PolarBase at Observatoire Midi Pyrénées, and observations collected at the European Southern Observatory, Chile (ESO programs 71.B-0617(A), 073.B-0074(A), and 085.B-0552(A)). Tables \ref{tab:basicdata_bulge}-\ref{tab:abundances_sn} are only available in electronic form at the CDS via anonymous ftp to cdsarc.u-strasbg.fr (130.79.128.5) or via \href{http://cdsweb.u-strasbg.fr/cgi-bin/qcat?J/A+A/}{http://cdsweb.u-strasbg.fr/cgi-bin/qcat?J/A+A/}}}

   \subtitle{IV. Zr, La, Ce, Eu}

   \author{R. Forsberg\inst{1} %orcID https://orcid.org/0000-0001-6079-8630
          \and H. Jönsson\inst{1,2}
          \and N. Ryde \inst{1,3}
             \and
             F. Matteucci \inst{4,5,6}
          %\fnmsep\thanks{Just to show the usage
          %of the elements in the author field}
          }
\institute{Lund Observatory, Department of Astronomy and Theoretical Physics, Lund University, Box 43, SE-221 00 Lund, Sweden\\ \email{rebecca@astro.lu.se}
\and
Materials Science and Applied Mathematics, Malm\"o University, SE-205 06 Malm\"o, Sweden
\and
Universit\'e C\^ote d'Azur, Observatoire de la C\^ote d'Azur, CNRS, Laboratoire Lagrange, Bd de l'Observatoire, CS 34229,
06304 Nice Cedex 4, France
\and
Dipartimento di Fisica, Sezione di Astronomia, Universit\`a di Trieste, via G.B. Tiepolo 11, I-34131, Trieste, Italy
\and
I.N.A.F. Osservatorio Astronomico di Trieste, via G.B. Tiepolo 11, I-34131, Trieste, Italy
\and 
I.N.F.N. Sezione di Trieste, via Valerio 2, 34134 Trieste, Italy
}

   \date{Received July 18, 2019; accepted September 18, 2019}

% \abstract{}{}{}{}{} 
% 5 {} token are mandatory
 
  \abstract
%context (optional)  
{Observations of the Galactic bulge point at a formation through secular evolution of the disk instead of gas dissipation and/or mergers, as previously believed. This would imply very similar chemistry in the disk and bulge. Some elements, like the $\alpha$-elements are well studied in the bulge, but others, like the neutron-capture elements, are much less well explored. Stellar mass and metallicity are factors that affect the neutron-capture process. Due to this, the enrichment of the ISM and the abundance of neutron-capture elements vary with time, making them suitable probes for the Galactic chemical evolution.}
%aims
{In this work we make a differential comparison of neutron-capture element abundances determined in the local disk(s) and the bulge, focusing on minimising possible systematic effects in the analysis, with the aim of finding possible differences/similarities between the populations.}
%method
{Abundances are determined for Zr, La, Ce and Eu in 45 bulge giants and 291 local disk giants, from high-resolution optical spectra. The abundances are determined by fitting synthetic spectra using the SME-code. The disk sample is separated into thin and thick disk components using a combination of abundances and kinematics.}
%results 
{We find flat Zr, La, Ce trends in the bulge, with a $\sim 0.1$ dex higher La abundance compared with the disk, possibly indicating a higher s-process contribution for La in the bulge. [Eu/Fe] decreases with increasing [Fe/H], with a plateau at around [Fe/H] $\sim -0.4$, pointing at similar enrichment as $\alpha$-elements in all populations.}
%conclusions
{We find that the r-process dominated the neutron-capture production at early times both in the disks and bulge. [La/Eu] for the bulge are systematically higher than the thick disk, pointing to either a) a different amount of SN II or b) a different contribution of the s-process in the two populations. Considering [(La+Ce)/Zr], the bulge and the thick disk follow each other closely, suggesting a similar ratio of high/low mass asymptotic giant branch-stars.}


   \maketitle
%
%________________________________________________________________

\section{Introduction}
Our view of the structure and formation of the Galactic bulge has changed dramatically over the past decade. Earlier, the prevailing view was that the bulge is a spheroid in a disk, formed in an early, rapid, dissipative collapse \citep[e.g.][]{immeli:04}, naturally resulting from, for instance, major mergers converting disks to classical bulges \citep[e.g.][]{shen:15}. However, with new findings and data accumulating, what we call the bulge is today predominately considered to be mainly the inner structures of the Galactic bar seen edge-on \citep[e.g.][]{portail:17}. The details of its structure and timescales for its formation is, however, not clear \citep[e.g.][]{barbuy:18}.

Metallicity distributions and abundance-ratio trends with metallicity provide important means to determine the evolution of stellar populations, also in the bulge. Trends of different element groups formed in different nucleosynthetic channels provide strong complementary constraints. Also, comparisons of trends between different stellar populations, e.g. the local thick disk, can constrain the history of the bulge. Whether or not there is an actual difference in abundance trends with metallicity between the bulge and the local thick disk is not clear \citep{MCwilliam16,barbuy:18,zasowski:19,lomaeva:19}. Some elements such as Sc, V, Cr, Co, Ni and Cu show differences in some investigations, whereas others show great similarities. New abundance studies minimising systematic uncertainties are clearly needed.

An important nucleosynthetic channel that has not yet been thoroughly investigated in the bulge is that of the heavy elements, namely the neutron-capture elements. These can be divided into two groups: the slow (s)- and rapid (r)-process elements, depending on the timescales between the subsequent $\beta$-decay and that of the interacting neutron flux \citep{burbidge1957}. The neutron flux in the s-process is such that the time scale of interaction is \textit{slower} than the subsequent $\beta$-decay, making the elements created in this process to trace the valley of stability, whilst it is the other way around for the r-process, resulting in creation of heavier elements. As a point of reference, the s-process therefore produces the lighter elements after iron (A $\geq$ 60), whereas the r-process is the dominating production process for the heaviest elements. Nonetheless, it is important to keep in mind that the production of heavier elements is a combination of the two processes and a ``s- or r-process element'' simply refers to an element having a dominating contribution from one of the processes. The neutron densities required for the s- and the r-process are $\leq 10^{7}-10^{15}$ cm$^{-3}$ \citep{busso1999,karakaslattanzio:2014} and somewhere between $10^{24} - 10^{28}$ cm$^{-3}$ \citep{kratz2007}, respectively, putting some constraints on the astrophysical sites where they can occur.

The s-process can in turn be divided into three sub-processes: the \textit{weak}, \textit{main} and \textit{strong} s-process, taking place in massive stars (weak) and asymptotic giant branch (AGB) stars (main, strong). Furthermore, the s-process \emph{elements} can be divided into the light, heavy and very heavy s-process elements, the naming originating from their atomic masses of A = 90, 138 and 208 (around Zr, La and Pb, respectively). A build-up is created at these stable nuclei (N = 50, 82 and 126, also known as magic numbers) due to isotopes with low neutron cross sections, creating bottlenecks in the production of heavier elements and in turn, peaks of stable isotopes. Thus, the naming first- second- and third-peak s-process is often used too for the light, heavy and very heavy s-process elements. In this work, light and heavy s-process elements produced in the main s-process will be analysed (Zr, La, Ce).

The main s-process takes place in the interior of low- and intermediate-mass AGB stars \citep{herwig:05,karakaslattanzio:2014} with the neutrons originating from the reactions ${}^{13}\text{C}(\alpha, n)^{16}\text{O}$ and ${}^{22}\text{Ne}(\alpha, n)^{25}\text{Mg}$. The second reaction takes place at higher temperatures in AGB stars with initial mass of $>4$ M$_{\odot}$. The process takes place in the so-called ${}^{13}\text{C}$-pocket in-between the hydrogen and helium burning shells during the third dredge-up \citep[TDU;][]{2017ApJ...835...97B}. Since AGB stars have an onset delay on cosmic scales, a non-negligible fraction of the s-process-dominated elements is likely to originate from the r-process at early times. Furthermore, the light s-process elements (first-peak s-process) can have a possible production from the weak s-process, taking place in helium core burning and in the subsequent convective carbon burning shell phase, in massive stars \citep{couch:1974}. However, previous observations can not, to a full extent, explain the abundance of the light s-process elements at early times and other possibilities of their origin have therefore been proposed \citep[e.g. LEPP;][]{travaglio2004,cristallo:15}.

The production site(s) for r-process elements is yet to be constrained, but the proposed sites are various neutron-rich (violent) events, such as core collapse supernovae (CC SNe), collapsars and the mergers of heavy bodies in binaries, like neutron star mergers \citep{sneden:2000,THIELEMANN2011346,2017ARNPS..67..253T}. The electromagnetic counterpart to the observed neutron merger GW170817 \citep{gw170817} indeed showed r-process elements. Research is still ongoing to determine whether or not neutron star mergers is the only, or even the dominating, source of r-process elements \citep[e.g.][]{2018SSRv..214...62T,cote:2018,siegel2019,kajino:2019}.

In order to put constraints on the neutron capture yields, it is important to have reliable observational abundances to compare with the models. In the review paper on the chemical evolution of the bulge by \cite{MCwilliam16}, the necessity of having properly measured abundances for the disk in order to have a reference sample for bulge measurements is stressed, which is provided in this work.

Regarding the determination of neutron-capture elements in bulge stars, such analyses have been made previously by \citet{johnsonetal2012}, \citet{vanderswaelmen}, and \citet{duong:19}. \citet{johnsonetal2012} studied stars in Plaut's field ($b = -8^{\circ}$) observed with the Hydra multifiber spectrograph at the Blanco 4m telescope, determining the abundances of Zr, La, Nd and Eu. Their [La/Fe] trend versus metallicity of the stars in the bulge field is clearly different from that of the thick disk. They thus conclude that the metal-poor bulge, or the inner disk, is likely chemically different from that of the thick disk. \citet{vanderswaelmen} studied Ba, La, Ce, Nd, and Eu in 56 Galactic bulge giants, observed with FLAMES/UVES at the VLT, finding that the s-process elements Ba, La, Ce, Nd, have decreasing [Ba,La,Ce,Nd/Fe] abundances with increasing metallicity, separating them from the flatter thick disk trends. Additionally, in the work by \citet{duong:19}, Zr, La, Ce, Nd, and Eu is measured for a large bulge sample at latitudes of $b = -10^{\circ}$, $-7.5^{\circ}$ and $-5^{\circ}$, observed with the HERMES spectrograph on the Anglo-Australian Telescope. They find indications of the bulge having a higher star formation rate than that of the disk.

\citet{johnsonetal2012} and \citet{vanderswaelmen} compare their bulge abundances with previously determined disk abundances, mainly from dwarf stars, which might obstruct the interpretation of the comparative abundances due to the risk of systematic uncertainties between analyses of dwarf and giant stars\footnote{\citet{duong:19} do, to a large extent as possible, use the same atomic data and analysis method in their work as their comparison sample, GALAH \citep{buder:18}, to minimise systematic offsets.}. Previous works by \citet{melendez:2008} and \citet{gibs:II} stressed the importance of comparing stars within the same evolutionary stage. Furthermore, in investigations of atomic diffusion and mixing in stars \citep{korn:07,lind:08,nordlander:2012,gruyters:2016,souto:2019,liu:2019}, it has been shown that dwarf stars might have systematically lower elemental abundances compared to evolved stars, suggesting that abundances measured from dwarf stars are too low. The magnitude of this depletion is measurable and should, in general, be considered for the relevant elements in order to properly probe the Galactic composition and its evolution based on dwarf stars. 

In this paper, we study the four neutron-capture elements Zr, La, Ce, and Eu, determined from optical spectra of giants observed with FLAMES/UVES for the bulge sample. We compare the obtained abundance-ratio trends with that of the local disk, obtained from a comparison sample of similarly analysed giants (observed with FIES at high resolution in the same wavelength range). Section 2 describe the bulge and disk samples. The same methodology for determining the stellar parameters and abundances (a carefully chosen set of spectral lines) ensures a minimisation of the systematic uncertainties in the comparison of the two samples, following the same methodology as the previous papers in this series; \citet{Jonsson2017a,Jonsson2017b,lomaeva:19}, see Sect. \ref{sect: analysis}. We present the results in Sect. \ref{sect: results} and discuss these in Sect. \ref{sect: discussion}.

\section{Observations}
\label{sect: observations}
\begin{figure*}
   \centering
\includegraphics[width=\hsize]{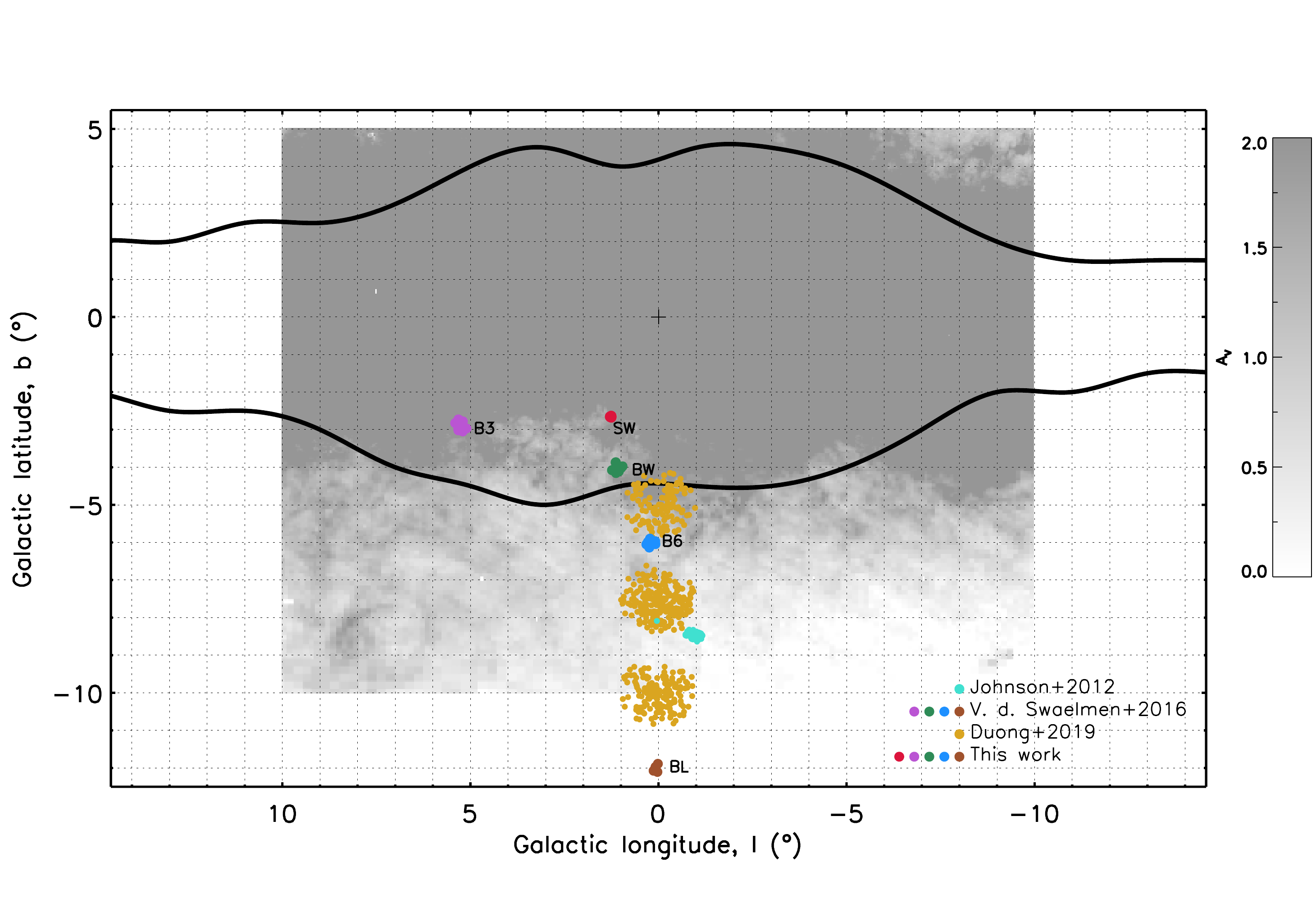}
 \caption{Map of the Galactic bulge showing the five analysed fields (SW, B3, BW, B6, and BL). The bulge samples from \cite{johnsonetal2012}, \cite{vanderswaelmen} and \citet{duong:19} are also marked in the figure. The dust extinction towards the bulge is taken from \citet{gonzalez:11,gonzalez2012} scaled to optical extinction \citep{cardelli}. The scale saturates at A$_{\text{V}} = 2$, which is the upper limit in the figure. The COBE/DIRBE contours of the Galactic bulge, in black, are from \citet{weiland:94}.}
\label{fig: bulge fields}
    \end{figure*}

\subsection{Bulge sample}
Since large amounts of dust lies in the line-of-sight toward the Galactic centre, resulting in a high optical extinction, observing bulge stars can be challenging at optical wavelengths.  %(normally IR observations are used instead). 
Our ambition was to include fields as close to the centre of the bulge as possible, whilst keeping to regions where the extinction is manageable.  %As a result, it was only possible to observe stars in the outer part of the bulge, where the extinction is less severe. %urthermore, determining the stellar parameters for these observations is still difficult (e.g. Rich et al. 2012), but improvements are an ongoing subject of research (e.g. Ryde et al. 2016; Schultheis et al.2016; Rich et al. 2017; Nandakumar et al. 2018).

The Galactic bulge sample consists of 45 giants (see Table \ref{tab:basicdata_bulge}). The spectra were obtained using the spectrometer FLAMES/UVES mounted on the VLT, Chile, observed in May-August 2003-2004. 27 of these spectra were also analysed in \citet{vanderswaelmen}. In addition to these, 18 spectra from the Sagittarius Window, ($l,b)=(1.29^\circ,-2.65^\circ$), lying closer to the Galactic plane in a region with relatively low extinction, is added to the sample analysed here. These were observed in August 2011 (ESO programme 085.B-0552(A)). In total, five bulge fields are included in the bulge sample: SW (the Sagittarius Window), BW (Baade's Window), BL (the Blanco field), B3 and B6\footnote{The naming of the fields follows the convention seen in \citet{lecureur:07}}. The fields can be seen in Fig. \ref{fig: bulge fields}, overlaid on an optical extinction map, together with the fields analysed in \citet{johnsonetal2012} and \citet{duong:19}. From Fig. \ref{fig: bulge fields} one can see that the SW field lies in a region of relatively low extinction and closer to the Galactic plane then the other fields. 

%Although the definition of the bulge field is a bit vague, using the \citet{barbuy:18} definition of the bulge as the region on the sky within $b=\pm10^{\circ}$ and $l=\pm10^{\circ}$, the Blanco field is on the verge of being a bulge field. Even so, we chose to include these spectra in our bulge compilation. In Figure \ref{fig: bulge fields} the fields used in \cite{johnsonetal2012} and \citet{vanderswaelmen} are marked out too, where we have 27 stars overlapping with that of \cite{vanderswaelmen}.

The FLAMES/UVES instrument allows for simultaneous observation of up to seven stars. Depending on the extinction, and local conditions, each setting in our observations required an integration time of somewhere between 5–12 hours. The achieved signal-to-noise ratios (S/N) of the recorded bulge spectra are between 10-80. The resolving power of the spectra is R $\sim$ 47 000 and the usable wavelength coverage is limited to the range 5800 and 6800 Å. 

The distances to our bulge stars are estimated to range between 4-12 kpc from the Solar System \citep{bailer-jones:2018}, placing the stars within the Galactic regions classified as the bulge by \cite{wegg:2015}. Although it should be noted that distance estimation can be rather troublesome and Gaia DR2 \citep{gaiacollab:2016,gaiacollab:2018} reports a parallax uncertainty higher than 20 \% for a majority of our bulge stars.

\subsection{Disk sample}
The disk sample consists of 291 giants stars, a majority of these placed within 2 kpc from the Solar System (see Table \ref{tab:basicdata_sn}). The bulk of the sample is observed at the Nordic Optical Telescope (NOT), La Palma, using the FIbre-fed Echelle Spectrograph (FIES; \citet{telting:2014}), under the programme 51-018 (150 stars) in May–June 2015 and 53-002 (63 stars) in June 2016. 41 spectra were taken from the stellar sample in \citet{thygesen:2012}, also observed using the FIES at the NOT. An additional 18 spectra were downloaded from the FIES archive. Lastly, 19 spectra were taken from the PolarBase data base \citep{petit:2014} where NARVAL and ESPaDOnS have been used (mounted on Telescope Bernard Lyot and Canada-France-Hawaii Telescope, respectively). FIES and PolarBase have similar resolving powers of R $\sim 67 000$ and R $\sim 65 000$, respectively.

All three spectrometers cover wide regions in the optical domain, but in order to maximise the coherency in this work, the wavelength region used is restricted to that of the the bulge spectra; 5800-6800 Å. The resulting S/N of the FIES spectra are around 80-120 per data point in the reduced spectrum. Similar values can be found for the PolarBase spectra whereas the \citet{thygesen:2012} spectra have a lower S/N of about 30–50. Details about how the S/N was calculated can be found in \citet{Jonsson2017a}.

The reduction of the FIES spectra was preformed using the standard FIES pipeline and the \citet{thygesen:2012} and PolarBase data was already reduced. A crude normalisation of all spectra was done initially with the IRAF task {\tt continuum}. Later in the analysis, the continuum is re-normalised more carefully by a manual placement of continuum regions and subsequently fitting a straight line to these, allowing a higher precision of the abundance determination (more on this in Sect. \ref{section: abundance determination}).

Telluric lines have not been removed from the spectra, instead, a telluric spectrum from the Arcturus atlas \citep{2000IAUJD...1E..26H} has been plotted over the appropriately-shifted observed spectra and affected regions have been avoided on a star-by-star basis.

%%%%%%%%%%%%%%%%%%%%%%%%%%%%%%%%%%%%%%%%%%%%%%%%%%%%%%%%%%%%%%%%%%%%%%%%%%%%%%%%%%%%%%%

\section{Analysis}
\label{sect: analysis}
The analysis of the spectra and the determination of the stellar abundances follows the same methodology as described in the previous papers in this series; \citet{Jonsson2017a,Jonsson2017b} and \citet{lomaeva:19}. This section describes the general methodology as well as the specifics relevant for this work.

\subsection{General methodology}
\begin{figure}
   \centering
\includegraphics[width=\hsize]{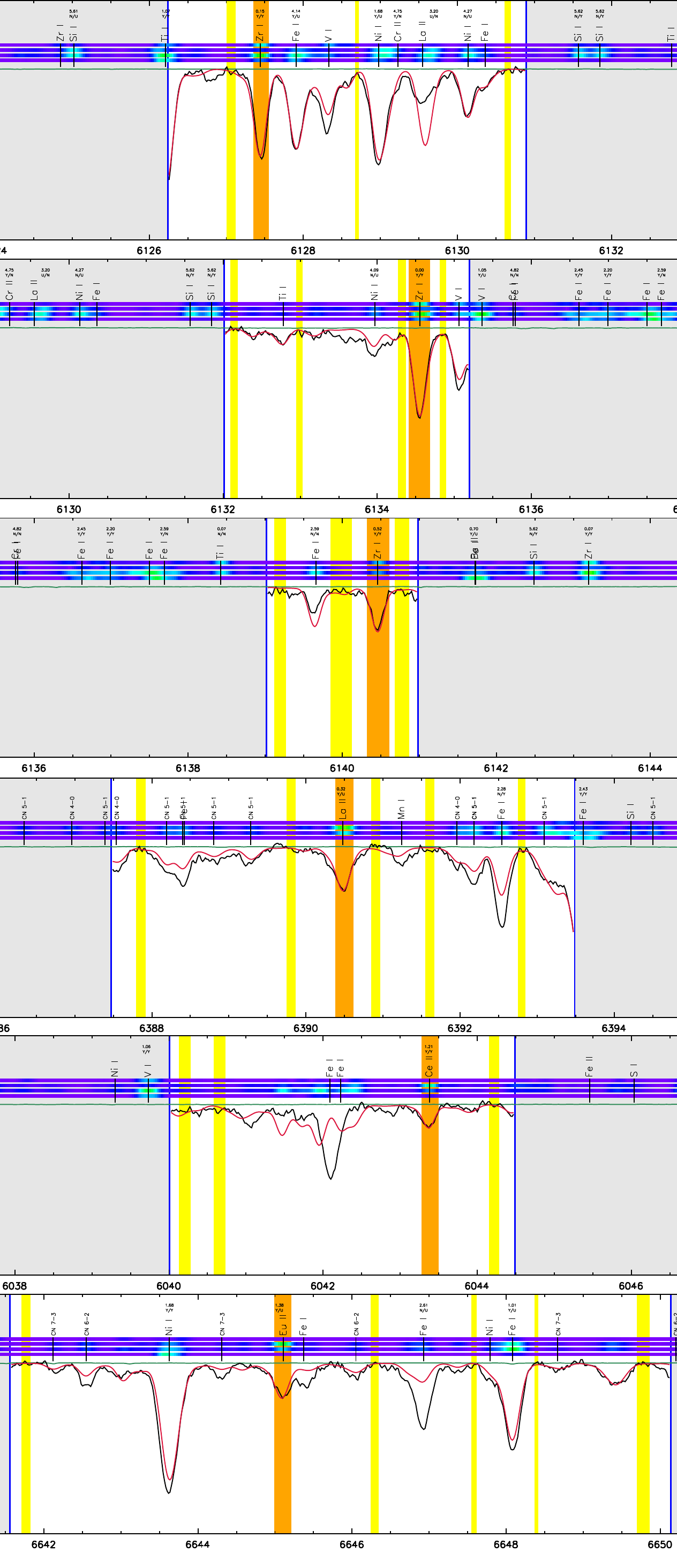}
 \caption{The observed spectrum (black) of the bulge star B6-F1 (S/N = 54). The lines for abundance determination of Zr (three lines), La, Ce and Eu (one line each) are marked out as the orange regions. The yellow regions are the manually placed continuum and the red spectrum is the synthetic one. The segments within which the synthetic spectrum is modelled are marked as the white wavelength regions between the blue vertical lines in each panel. The four horizontal lines above each spectrum indicates the lines' sensitivity in the stellar parameters T$_{\text{eff}}$, $\log g$, [Fe/H] and $\xi_{\text{micro}}$, respectively, where green is more sensitive than blue.}
 \label{fig: spectrum bulge high S/N}%
\end{figure}

To determine the stellar abundances synthetic spectra are modelled using the tool Spectroscopy Made Easy \citep[SME, ][]{sme,SME2017}. For a given set of stellar parameters (T$_\mathrm{eff}$, $\log g$, [Fe/H], and microturbulence, $\xi_\mathrm{micro}$), SME interpolates in a grid of pre-calculated model atmospheres and calculates a synthetic spectrum of a region of choice. By defining line and continuum masks over spectral regions of interest, SME can simultaneously fit, using $\chi^2$-minimisation \citep{marquardt1963algorithm}, both stellar photospheric parameters and/or stellar abundances. Figure \ref{fig: spectrum bulge high S/N} shows the line definitions and continuum placements for the bulge star B6-F1 and the spectral lines used in the analysis. %, as well as Figure \ref{fig: spectrum bulge low S/N}, \ref{fig: arcturus spectrum disk} 

The stellar parameters of the stars analysed are determined as described in Sect. \ref{section: stellar params} below. Metallicity-scaled solar abundances \citep{solar:sme} are assumed in SME, except for the $\alpha$-elements that have already been determined in \citet{Jonsson2017b}. 

SME uses a grid of MARCS models\footnote{Available at \url{marcs.astro.uu.se}} \citep{marcs:08} that adopts spherical symmetry for $\log g < 3.5$, which is the case for a majority of our stars, otherwise plane parallel. Some non-local thermodynamic equilibrium (NLTE) effect has been reported for the elements analysed here: Zr is shown by \citet{2010AstL...36..664V} to be weakly dependent on temperature; and \citet{2000A&A...364..249M} find that they need small NLTE corrections of the order of +0.03 dex for Eu in their analysis of cool dwarfs. Nonetheless, the analysis in this work is done under the assumption of LTE. 

\subsection{Stellar parameters}
\label{section: stellar params}
The stellar parameters used are determined in \citet{Jonsson2017a,Jonsson2017b} (where a more detailed description can be found) by fitting synthetic spectra for unsaturated and unblended Fe I and Fe II lines, Ca I lines and $\log g$ sensitive Ca I line wings, while T$_{\text{eff}}$, $\log g$, [Fe/H], $\xi_\mathrm{micro}$, and [Ca/Fe] were set as free parameters in SME. Fe I has NLTE corrections adopted from \citet{lind2012:nlte}. The reported uncertainties for these parameters in \citet{Jonsson2017a,Jonsson2017b} for a typical disk star of S/N $\sim 100$ are T$_{\text{eff}}$ $\pm$50 K, $\log g$ $\pm$ 0.15 dex, [Fe/H] $\pm$ 0.05 dex and $\pm$0.1 km/s for $\xi_\mathrm{micro}$. For a typical bulge star, the S/N is significantly lower (median of 38), and hence the uncertainties greater; T$_{\text{eff}} \pm 100$ K, $\log g \pm 0.30$~dex, [Fe/H] $\pm 0.10$~dex and $\xi_{\text{micro}} \pm$ 0.2 km\,s$^{-1}$. These values are later used in the uncertainties estimations, see Sect. \ref{section:uncertainties}.

\subsection{Abundance determination}
\label{section: abundance determination}
The atomic line data used for the abundance determination are collected from the \textit{Gaia}-ESO line list version 6 (\citealt{heiter15}, Heiter et al., in prep.). From here we get wavelengths, excitation energies and transition probabilities (as well as broadening parameters, when existing). The transition probabilities for the elements investigated here, Zr, La, Ce, and Eu, come from \cite{BGHL}, \cite{LBS}, \cite{LSCI}, and \cite{LWHS}, respectively.  All available lines for these elements in the given wavelength region (5800-6800 Å) were investigated individually in order to exclude lines that could not be modelled properly (due to blends, bad atomic data or other systematics). As for Zr, where three separate lines were suitable for abundance determination, the lines were ultimately fitted simultaneously. Finally, the determined SME abundances were, in the post-process, re-normalised to the most up-to-date solar values provided by \citet{grevesse:2015}. The final set of lines used for abundance determination is presented in Table \ref{table: final line list}. Apart from the atomic lines, we include the molecules C$_2$ \citep{2013JQSRT.124...11B} and CN \citep{2014ApJS..214...26S} in the synthesis.

\begin{table}
\centering
\caption{\label{table: final line list}Atomic lines used in the analysis. The elements and ionisation stages are given in Col. 1, the transition wavelengths in Col. 2, and the $\log(gf)$ values are listed in Col. 3. The excitation energies of the transitions lower level are given in Col. 4. The $\log(gf)$ data included in the Gaia-ESO line lists comes from \cite{BGHL} (Zr), \cite{LBS} (La), \cite{LSCI} (Ce) and \cite{LWHS} (Eu).}
\centering
\centerline{
\begin{tabular}{l c c c}
Element & Wavelength [Å]  & $\log(gf)$ & $\chi_{\text{ exc}}^{\text{ low}}$ [eV]\\ \hline \hline
Zr I &  6127.440 &  -1.06 & 0.15 \\
Zr I &  6134.550 &  -1.28 & 0.00 \\
Zr I &  6140.460 &  -1.41 &  0.51\\ \hline 
La II &  6390.457 &  -2.01 &  0.32 \\ 
La II &  6390.469 & -2.08 &  0.32 \\
La II &  6390.486 & -1.90 &  0.32  \\
La II &  6390.501 &  -2.08 & 0.32  \\ \hline 
Ce II &  6043.373 &  -0.48 &  1.21 \\ \hline 
Eu II & 6645.057 &  -0.84 &  1.38  \\ 
Eu II & 6645.060 &  -0.78 &  1.38   \\ 
Eu II & 6645.068 &  -2.13 &  1.38   \\ 
Eu II & 6645.074 &  -0.84 &  1.38  \\ 
Eu II & 6645.083 &  -0.91 &  1.38   \\
Eu II & 6645.086 &  -0.90 &  1.38  \\
Eu II & 6645.098 &  -0.60 &  1.38  \\ 
Eu II & 6645.101 &  -0.95 &  1.38  \\ 
Eu II & 6645.121 & -1.01 &  1.38   \\ 
Eu II & 6645.137 &  -1.09 &  1.38  \\ 
Eu II & 6645.149 &  -1.19  &  1.38  \\
\end{tabular}}
\end{table}

\begin{table*}
\centering
\caption{Isotope information of the elements. Col. 2 gives the baryon number of the stable isotopes that contribute to at least 1 \% to the solar system abundance. Col. 3 gives the corresponding relative isotopic abundances of the stable isotopes as measured in the Sun, with references in the last column.}
\label{table: isotopic information table}
\begin{tabular}{l|l|l|l}
Element(Z) & Baryon number  & Relative abundance & Reference  \\ \hline \hline 
Zr(40)    & 90:91:92:94:96     & 51:11:17:17:3      & \cite{1983IJMSI..50..219N}\\
La(57)     & 139      & 100  & \cite{2005IJMSp.244...91D}\\
Ce(58)   & 140:142    & 88:11      & \cite{1995IJMSI.142..125C}\\
Eu(63)    & 151:153      & 48:52     & \cite{1994IJMSI.139...95C}\\
\end{tabular}
\end{table*}

For La and Eu hyperfine splitting (hfs) had to be taken into account. By not taking hfs into account there is a risk of overestimating the measured abundance \citep{2000ApJ...537L..57P,thorsbro:18}. Additionally, isotopic shift (IS) has to be considered for Zr, Ce and Eu. The shift is caused by the isotopes having shifted energy levels, resulting in radiative transitions with shifted wavelengths. IS is included by manually identifying the set of transitions for each isotope in the line list and scaling the $\log(gf)$ to the relative solar isotopic abundances, see Table \ref{table: isotopic information table}.

\subsection{Population separation}
\label{section: population sep}
The classification of the stellar populations in the disk (thin/thick) can be done in several ways, by kinematics, age, geometry and chemistry. Even so, the separation of these two components is somewhat debated and the transition between them might be a gradient rather than a clear separation. The results by \citet{hayden:15} show that the scale length of the thin disk extends further out than that of the thick disk. The thick disk has been shown to be enriched in $\alpha$-elements, compared to that of the thin disk, in addition to thick disk stars having higher total velocities whilst slower rotational velocities \citep{bensby:14}. 

In \citet{lomaeva:19} the separation into the two populations is computed for our disk sample, using a combination of stellar metallicity, abundances ([Ti/Fe] as determined in \citet{Jonsson2017b}) and kinematics. The radial velocities from Table \ref{tab:basicdata_sn}, proper motions from Gaia DR2 \citet{gaiacollab:2016,gaiacollab:2018} and distances from \citet{mcmillan:2018distances} are used to calculate the total velocities\footnote{$V_{\text{tot}}^2 = U^2 + V^2 + W^2$}. In total, kinematic data were available for 268 stars in the disk sample. The clustering method \texttt{Gaussian Mixture Model} (GMM), obtained from the \texttt{scikit-learn} module for Python \citep{pedregosa2011scikit}, is used to cluster the disk data into the two components. The reader is referred to \citet{lomaeva:19} for more details.
%The GMM assumes that the sample consists of a weighted sum of Gaussian sub-distributions and uses an iterative Expectation-Maximisation (EM) algorithm to obtain the GMM parameters (mean vectors, covariance matrices and mixture weights). The algorithm calculates the probability of the given data to belong to one of the clusters\footnote{The amount of clusters must be defined in GMM, in this case two.} and updates the parameters of the Gaussian sub-distributions using the estimated probabilities. This is done up until the point that the fit to the data converges. 

%%%%%%%%%%%%%%%%%%%%%%%%%%%%%%%%%%%%%%%%%%%%%%%%%%%%%%%%%%%

  \begin{figure*}
   \centering
\includegraphics[width=\hsize]{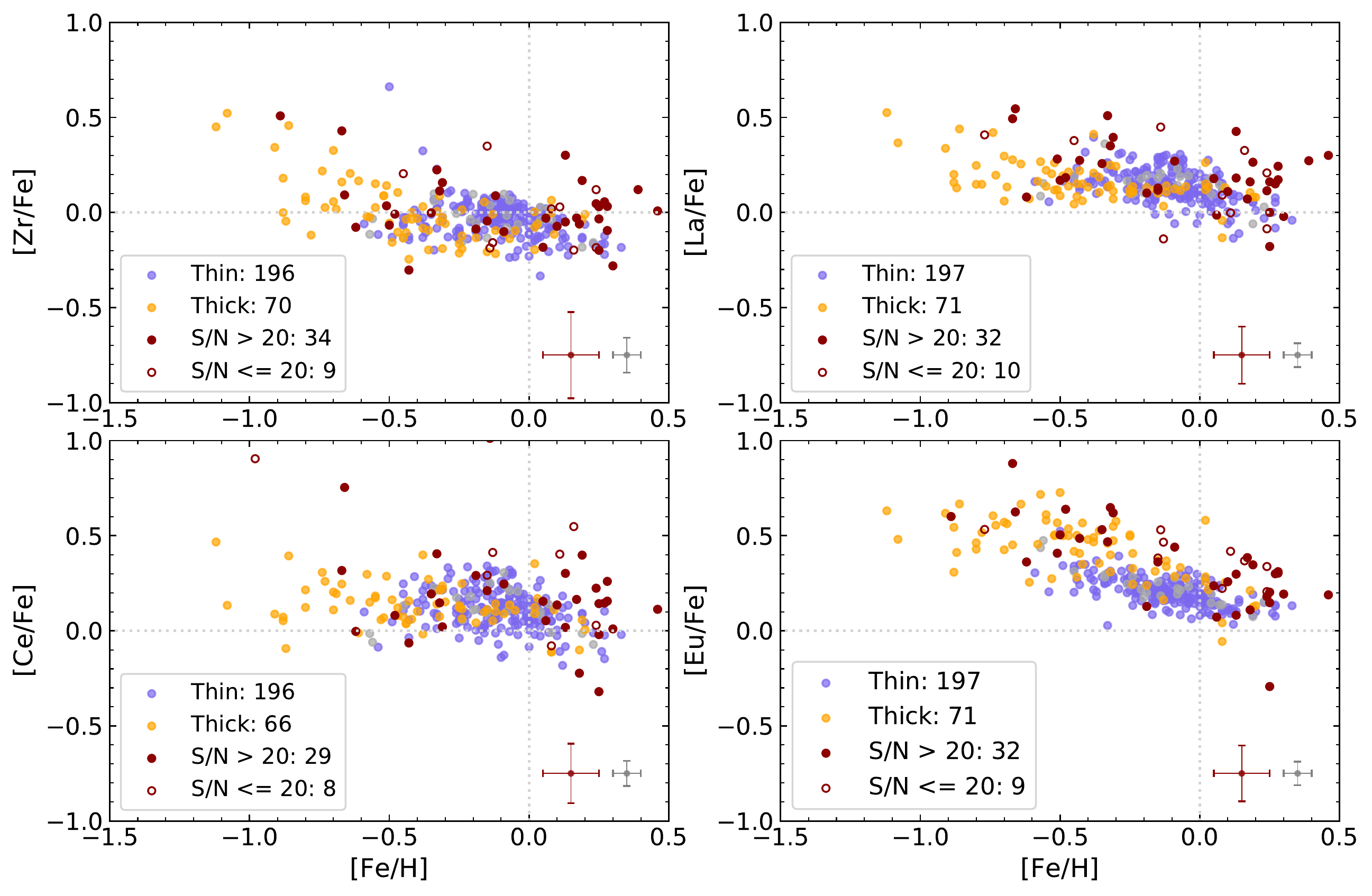}
 \caption{The abundance ratio trends with metallicity, [X/Fe] against [Fe/H], for the thin- (blue) and thick-disk (yellow) stars as well as the bulge stars (red). Since not all abundances were possible to determine in all spectra, the number of stars in each sample is included in the legend. Filled dark red indicate bulge stars with a S/N above 20, whereas the hollow red circles indicate a S/N equal to or less than 20. Some of the disk stars could not be classified as thick or thin disk stars; these are marked as grey dots. The typical uncertainty for the disk and the bulge sample, as described in Sect. \ref{section:uncertainties}, is marked in the lower right corner of every plot.}
              \label{fig: disk comparative}%
    \end{figure*}

\section{Results}
\label{sect: results}
Our derived abundance ratios, [X/Fe], for Zr, La, Ce, and Eu, are plotted against [Fe/H] in Fig. \ref{fig: disk comparative}. The population separation is applied to the disk sample and the number of stars in each population for which we could determine the abundance in question is noted in every panel. The bulge sample is plotted on top of the disk trends, differentiating between spectra of higher or lower S/N-ratio of 20. The typical uncertainty is noted in the plots, where the estimation of these is described in Sect. \ref{section:uncertainties}.

\subsection{Uncertainties}
\label{section:uncertainties}
Systematic errors generally originate from incorrectly determined stellar parameters, model atmosphere assumptions, continuum placement and atomic data. This makes these errors hard to estimate. To get a sense of the systematic uncertainties, one can compare to reference stars. In \citet{Jonsson2017a} they compare the determined stellar parameters to those of three overlapping Gaia benchmark stars determined in \citet{jofre:2015} and find that these are within the uncertainties of the Gaia benchmark parameters.
%, except for the surface gravity and microturbulence of $\mu$Leo that is off a little bit. However, this amount will hardly affect the model atmospheric structure, in addition only weak lines are used for abundance determination, where $v_{\text{mic}}$ has a very small influence. The otherwise excellent agreement for these three references thus suggest that there are no clear systematic uncertainties in the stellar parameters.

%\begin{table}
%\centering
%\label{table: gaia benchmark stellar param comp} 
%\caption{Stellar parameters of overlapping Gaia benchmark stars: $\alpha$Boo (HIP69673) 
%$\mu$Leo (HIP48455) and $\beta$Gem (HIP37826). Top row: the results presented in \cite{jofre:2015}; Bottom row: the stellar parameters used in this work.}
%\begin{tabular}{c|c|c|c|c}
%Star    & T$_\text{eff}$ {[}K{]} & $\log g$ {[}dex{]} & {[}Fe/H{]} {[}dex{]} & $v_\text{mic}$ [km/s]    \\ \hline \hline
%$\alpha$Boo &   4286$\pm$35 & 1.64$\pm$0.09  & -0.57$\pm$0.08 & 1.58$\pm$0.12  \\
%  &   4251    &  1.72    &   -0.57    &   1.64      \\ \hline
%$\mu$Leo   &  4474$\pm$60  & 2.51$\pm$0.11 & 0.20$\pm$0.15  & 1.28$\pm$0.26 \\
%   & 4461    &  2.65  &  0.23   &   1.55   \\  \hline
%$\beta$Gem & 4858$\pm$60 & 2.90$\pm$0.08 & 0.08$\pm$0.16 & 1.28$\pm$0.21 \\
%& 4835 &  2.93 &  0.07 &  1.24
%\end{tabular}
%\end{table}

All spectra are analysed using the same line and continuum masks as well as the same atomic data, minimising possible random uncertainties. Therefore, the random uncertainties are to primarily be found in the (random) uncertainties of the stellar parameters. An approach to estimate the random uncertainties due to changes in the stellar parameters, is to analyse a typical spectrum several times using parameters that all vary within given distributions. The same method for estimating the uncertainties was used in \citet{lomaeva:19}. 

Hence, using the FIES spectrum of the standard star Arcturus\footnote{The giant star Arcturus (also known as $\alpha$-Boo or HIP69673) has been analysed extensively due to its brightness, being the fourth brightest in the night sky, and is suitable as a reference of a typical giant star.}, uncertainties were added to its initial stellar parameters, meaning that the stellar parameters were changed simultaneously, for a set of 500 runs with modified stellar parameters. A Gaussian distribution is used to generate the uncertainties, using the reported stellar parameter uncertainties as standard deviation (see Sect. \ref{section: stellar params}). In the uncertainty estimation of the bulge abundances, we have not degraded the FIES Arcturus spectrum (with a resolution of 67000) to match that of the bulge spectra (R of 47000), but separate tests have shown this slightly lower resolution to have a negligible effect on the determined abundance.

The abundance uncertainties coming from the uncertainties in the stellar parameters are then calculated as
\begin{equation}
    \label{eq: uncertanty equation}
    \sigma A_{\text{parameters}} = \sqrt{|\delta A_{T_{\text{eff}}}|^2 + |\delta A_{\log g}|^2 + |\delta A_{\text{[Fe/H]}}|^2 + |\delta A_{v_{\text{micro}}}|^2},
\end{equation}
where, for non-symmetrical abundance changes, the mean value is used in the squared sums. The resulting uncertainties can be seen in Table \ref{table: uncertanties}. %The resulting abundances\footnote{The abundances A(X) are calculated using $A(X) = \log_{10}\left(N_X/N_Y\right)_{\text{star}} = \text{[X/H]}_{\text{star}} + \log_{10}\left(N_X/N_Y\right)_{\odot}$.} and probability density functions of these for the disk and the bulge can be seen in Appendix \ref{appendix: uncertanties}.

\begin{table}
    \centering
\caption{The estimated typical uncertainties for the disk and bulge sample using a generated set of stellar parameters for the giant star $\alpha$-Boo.}
\label{table: uncertanties}
    \begin{tabular}{c|c c c c}
         $\sigma A_{\text{parameters}}$ &  Zr & La  & Ce & Eu \\ \hline \hline 
         disk [dex] & 0.09 & 0.06 & 0.07 & 0.06 \\ 
         bulge [dex] & 0.23 & 0.15 & 0.16 & 0.15 \\ 
    \end{tabular}
\end{table}

%%%%%%%%%%%%%%%%%%%%%%%%%%%%%%%%%%%%%%%%%%%%%%%%%%%%%%%%%%%%

\section{Discussion}
\label{sect: discussion}
In this section we elaborate on the results. First, we compare separately our abundance trends for the disks and bulge with previous literature studies in Sect. \ref{section: comp to previous}. Second, and this is the core of this investigation, in Sect. \ref{section: disk and bulge comp} we consider a more in-depth comparative analysis between our abundances for the bulge and disks populations, both determined in the same way. This is done to minimise the systematic uncertainties as much as possible. We then proceed in considering and discussing comparative abundance ratios such as [Eu/Mg], [Eu/La] and [second-peak s/first-peak s], also in Sect. \ref{section: disk and bulge comp} as well as \ref{section: first second s process peak}. 

To highlight features of the trend-plots, the running mean of the samples are calculated and plotted (with a $1 \sigma$ scatter). The number of data points in the running window are set to roughly 15 \% of the sample sizes (thin disk, thick disk, bulge). As a result, the running mean (and scatter) does not cover the whole trend range. For the bulge sample, only data points with S/N $> 20$ are included in the running mean. From here and henceforth, the running mean-trend is the one referred to when describing [X/Fe] or [X/Y] ratios (except for Sect. \ref{sect: discussion disk sample}).

\subsection{Comparison with selected literature trends}
\label{section: comp to previous}
\subsubsection{Disk sample}
\label{sect: discussion disk sample}

\begin{figure*}
   \centering
\includegraphics[width=\hsize]{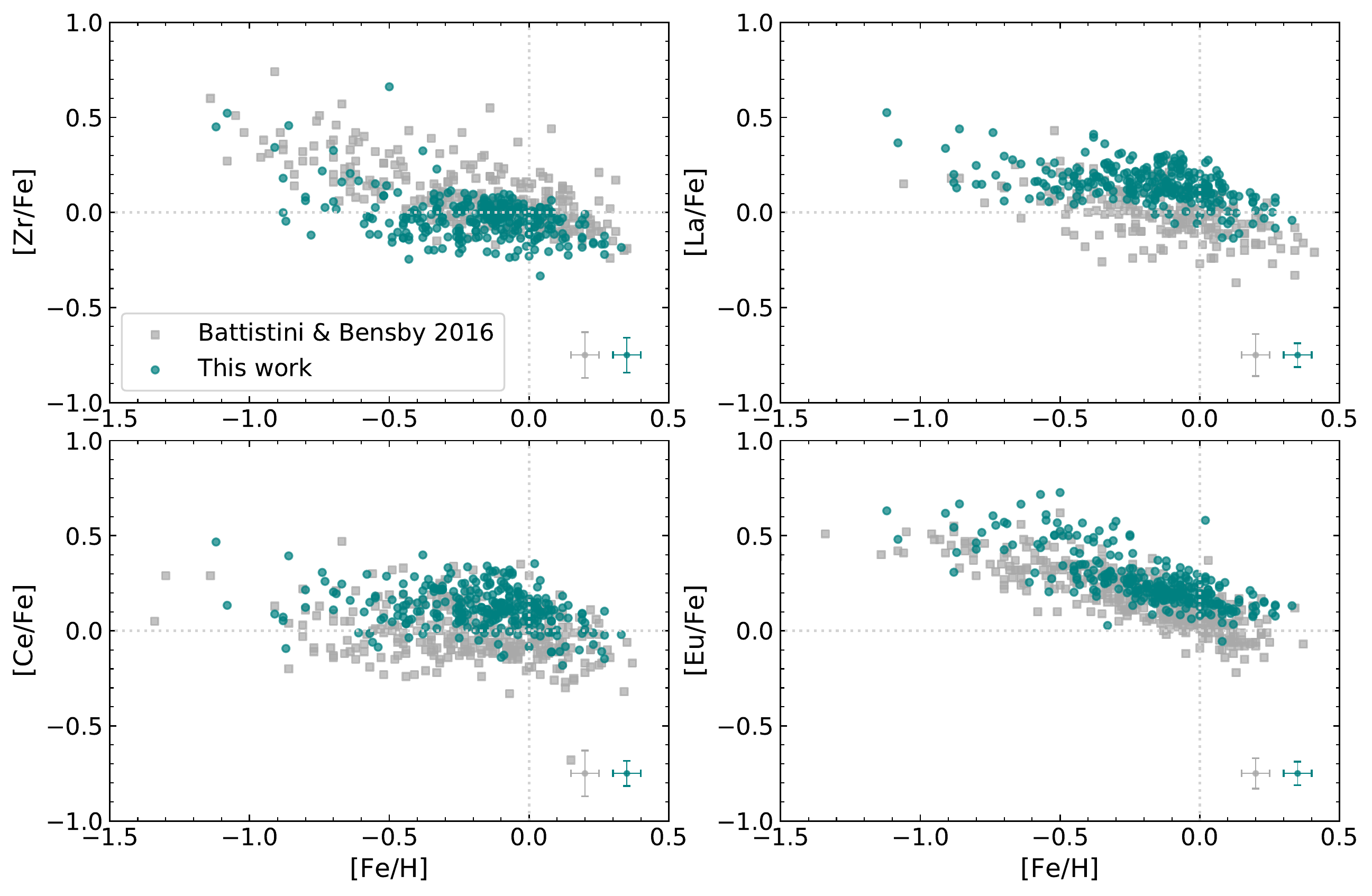}
 \caption{The determined disk abundances in this work (teal) compared with the determined abundances from \cite{battistinibensby:16} (grey). The typical uncertainty for both data sets are indicated in the lower right corner of every plot, where the uncertainties are taken from Table 6 in \citet{battistinibensby:16}.}
              \label{fig: disk BB16}%
\end{figure*}

\begin{figure*}
   \centering
\includegraphics[width=\hsize]{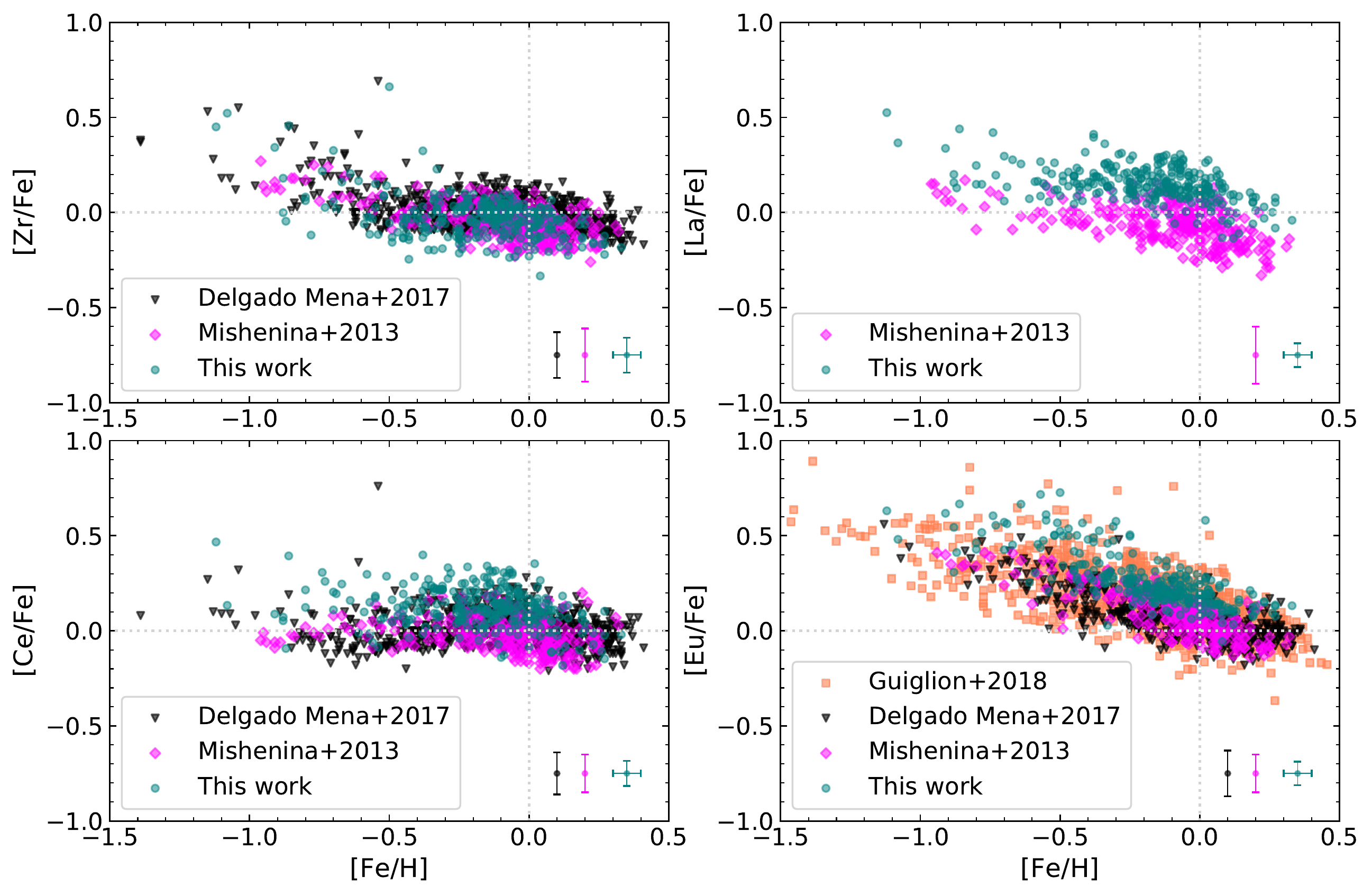}
 \caption{The determined disk abundances in this work (teal) compared with selected literature trends: \citet{Mishenina2013} (pink), \citet{DM17} (black) and \citet{Ambre18} (orange). The typical uncertainty, when available, are indicated in the lower right corner of every plot, where the uncertainties from in \citet{Mishenina2013} and \citet{DM17} (their Table 3 and Table 4, respectively) are for a low T$_{\text{eff}}$ star.}
              \label{fig: disk m13 DM17}%
\end{figure*}

In Fig. \ref{fig: disk BB16} we compare our determined disk abundances with those determined for dwarf stars in the disk by \citet{battistinibensby:16}. In general, the trends are similar for all elements, as well as the scatter in the determined abundance. The abundances of [La,Ce,Eu/Fe] seem to be systematically higher than those of \cite{battistinibensby:16} whereas the [Zr/Fe]-abundances seem a bit lower. The typical abundance uncertainties for \cite{battistinibensby:16} are 0.12, 0.11, 0.12 and 0.08 dex for Zr, La, Ce and Eu, respectively (their Table 6), which is somewhat higher than ours (see Table \ref{table: uncertanties}). The possible shifts in the abundances could be due to systematic differences in dwarf and giant stars or in differing atomic data such as using different lines in the abundance determination. Indeed, there is no overlap in the atomic lines used in these two data sets, except for the La line at 6390 Å, although \citet{battistinibensby:16} uses three additional lines for the La abundance determination.

Zr is a first-peak s-process element whereas La and Ce are second-peak s-process elements. [Zr,La/Fe] have somewhat decreasing abundances with increasing metallicities, with a flattening of abundances for [Fe/H] above $\sim -0.4$. The [Ce/Fe] trend is flatter than [Zr/Fe] and [La/Fe], explained by the higher s-process contribution in the Ce production (66 \%, 76 \% and 84 \% s-process contribution for Zr, La and Ce, respectively \citep{bisterzo:2014}).

The scatter for the [La/Fe] abundances is higher, $\sim 0.5$ dex, over the metallicity range [-0.2, 0], compared to the rest of the metallicity domain with $\sim 0.3$ dex. It indicates AGB stars producing the bulk of s-elements through the main s-process. The increase in scatter can most likely be explained by the mass range of AGB stars, which 1) enables stars to produce s-process elements at different metallicities (times) as well as 2) different amounts of production of the first-/second-peak s-process for different mass AGB stars (see Sect. \ref{section: first second s process peak}). The increasing abundances when [Fe/H] is below -0.5 for the s-process elements Zr and La point at a production by the r-process at early times (see [Eu/Fe]). 
In addition to \citet{battistinibensby:16}, our work compares with the abundances reported in \citet{Mishenina2013} (Zr, La, Ce, Eu) and \citet{DM17} (Zr, Ce, Eu), on dwarf stars in the local disk, see Fig. \ref{fig: disk m13 DM17}. The typical uncertainties from \citet{Mishenina2013} and \citet{DM17} are chosen from their estimates of low T$_{\text{eff}}$ stars, their table 3 and table 4, respectively.

For Eu, the trend decreases with increasing metallicities throughout our metallicity range, except for a plateau around [Fe/H] $< -0.6$. Eu has a reported r-process contribution of 94 \% \citep{bisterzo:2014} and the observed trend indicates that the r-process has a continuous enrichment in the Galaxy, similar to that of the $\alpha$-elements. Our Eu abundances compare well with those of \citet{Ambre18}, including some subgiant and giant stars in their sample, see Fig. \ref{fig: disk m13 DM17}. We note that our measurements, \cite{battistinibensby:16} and \cite{Ambre18} show slightly, on average, supersolar [Eu/Fe] abundances at solar metallicities, which is not seen in either \citet{Mishenina2013} nor \cite{DM17}. Of all the trends, ours is most systematically high not passing through the solar value at any metallicities.

\subsubsection{Bulge sample}
% v.d.Swaelmen: Same lines as for La, Ce and Eu. They do not include IS nor HFS (for EU).
% Johnson et al:
%Zr: 6134.545 - same, 6140.450 - same, 6143.200 - gave too scattered result.
%La: 6262.29 Å. Too blended with unknown lines (Y/N) GES
%Eu: same line with the same log gf components from \cite{LWHS}.
% Duong et al:
% Zr: not same (outside wavelength regime)
% La: not same (outside wavelength regime), besides 5805 which gave bad trend
% Ce: not same (outside wavelength regime)
% Eu: Same line, same atomic data.

\begin{figure*}
   \centering
\includegraphics[width=\hsize]{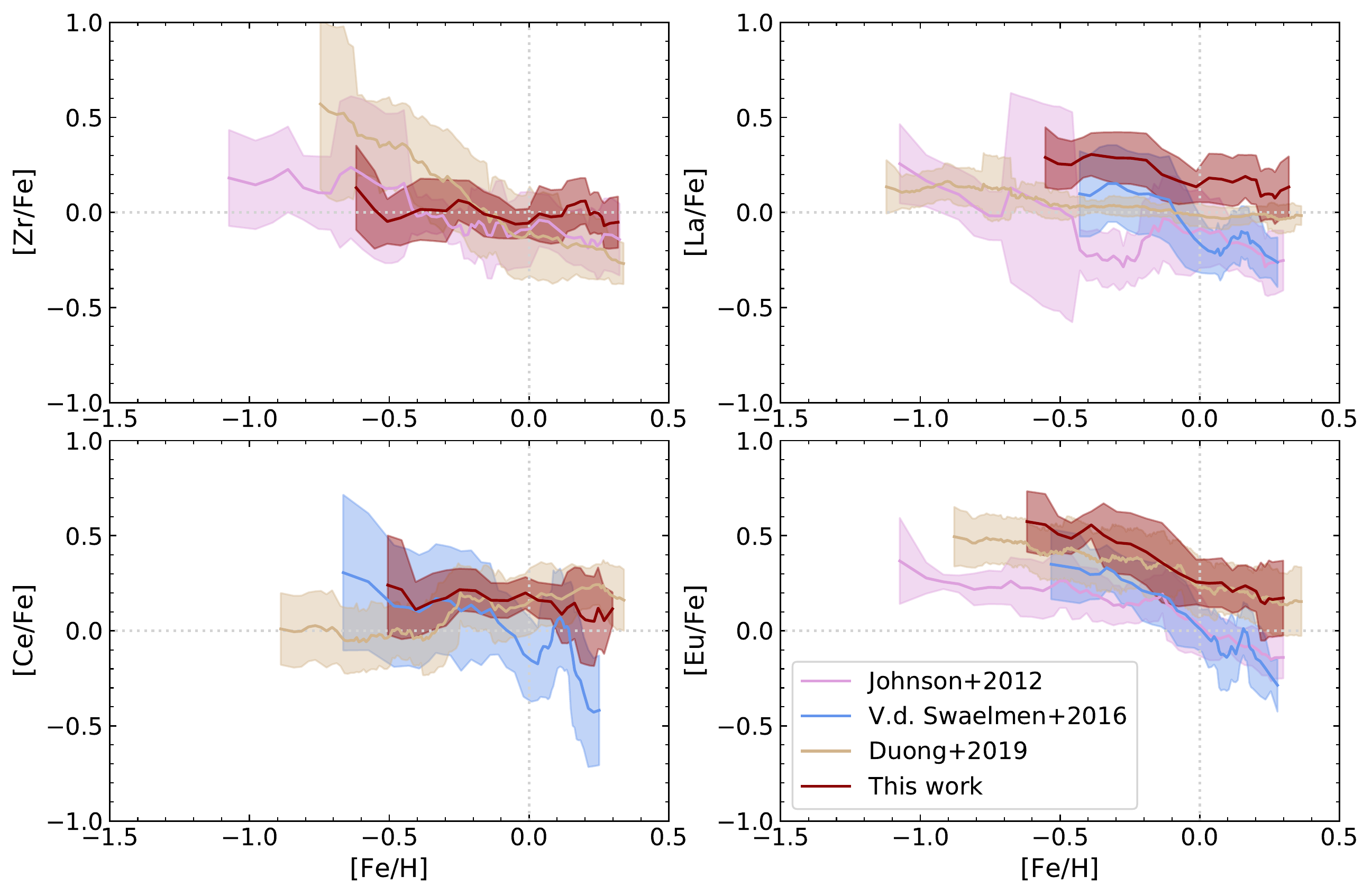}
 \caption{The running mean for the bulge abundances determined in this work (red solid line) compared with the calculated running mean based on the abundances in \citet{johnsonetal2012} (pink solid line),\cite{vanderswaelmen} (blue solid line) and \cite{duong:19} (beige solid line), with a 1$\sigma$ scatter (shaded regions, same colours as solid lines).}
              \label{fig: Running mean bulge comparative}%
\end{figure*}

In Fig. \ref{fig: Running mean bulge comparative} we compare our bulge trend with those observed in \citet{johnsonetal2012}, \citet{vanderswaelmen} and \citet{duong:19}. As for \citet{vanderswaelmen}, 27 of the stars/spectra overlap with their work and the same spectral lines are used for the abundance determination. Nonetheless, we observe different trends as well as measure Zr in these stars.

\textbf{Zirconium:} In general, our [Zr/Fe] trend with metallicity is flat, with an increase at lower metallicities [Fe/H] $< -0.5$. It should be noted that the running mean is rather poorly defined at the edges and the feature is based primarily on the two most metal-poor stars in Fig. \ref{fig: disk comparative}. Our trend agrees well with that of \citet{johnsonetal2012} within our overlapping metallicity ranges, whereas \citet{duong:19} has overall decreasing abundances with increasing metallicities. Above [Fe/H] $\sim$ 0.1, our [Zr/Fe] is solar while \citet{johnsonetal2012} and \citet{duong:19} have subsolar [Zr/Fe], ours pointing at a higher s-process contribution in the production of Zr.

\textbf{Lanthanum:} \citet{johnsonetal2012} reports a dip in [La/Fe] abundance around [Fe/H] $\sim -0.4$ which is not observed in either of the other studies, including ours. Both \citet{johnsonetal2012} and \citet{vanderswaelmen} produce decreasing [La/Fe] abundances with increasing metallicities, whilst both ours and \citet{duong:19} exhibit only a very small decrease of [La/Fe] with increasing [Fe/H]. In general our [La/Fe] abundances are higher than the other studies, which possibly could point at a higher s-process production in the bulge, compared to previous work. However, we note that our bulge abundances are expected to, similarly as the disk abundances, suffer from a systematic offset in the determined [La/Fe] abundance ratios, preventing us from making a firm claim.

\textbf{Cerium:} Our [Ce/Fe] trend is flat throughout our metallicity range. \citet{duong:19} also find a flat trend at solar scaled values, but with a slight step-wise increase at [Fe/H] $\sim -0.3$, thereafter following our trend. \citet{vanderswaelmen} find a different [Ce/Fe] trend with decreasing [Ce/Fe] values with increasing metallicities.

\textbf{Europium:} All the published [Eu/Fe] bulge trends and ours decrease with increasing metallicity, although with slightly different slopes and different offsets. The \citet{johnsonetal2012} study ranges to the lowest metallicities of all the samples. The trend of \citet{duong:19} and ours trace each other closely with super-solar abundances at all metallicities. The \citet{johnsonetal2012} and \citet{vanderswaelmen} trends follow each other well, in their overlapping metallicity region, with subsolar abundances above solar metallicities. There is an observable `knee' in the trend around [Fe/H] $\sim -0.4$, seen in all four works. Similarly as [La/Fe], our [Eu/Fe] abundances are higher than previous works, although due to the possible systematic offsets we cannot draw any firm conclusions from this. However, since the main purpose of this work is to make a differential analysis between the disk and bulge abundances in this work, the possible systematic offset in our analysis is of less importance.

\subsection{Disk and bulge comparison of the current study}  
\label{section: disk and bulge comp}

\begin{figure*}[ht!]
   \centering
\includegraphics[width=\hsize]{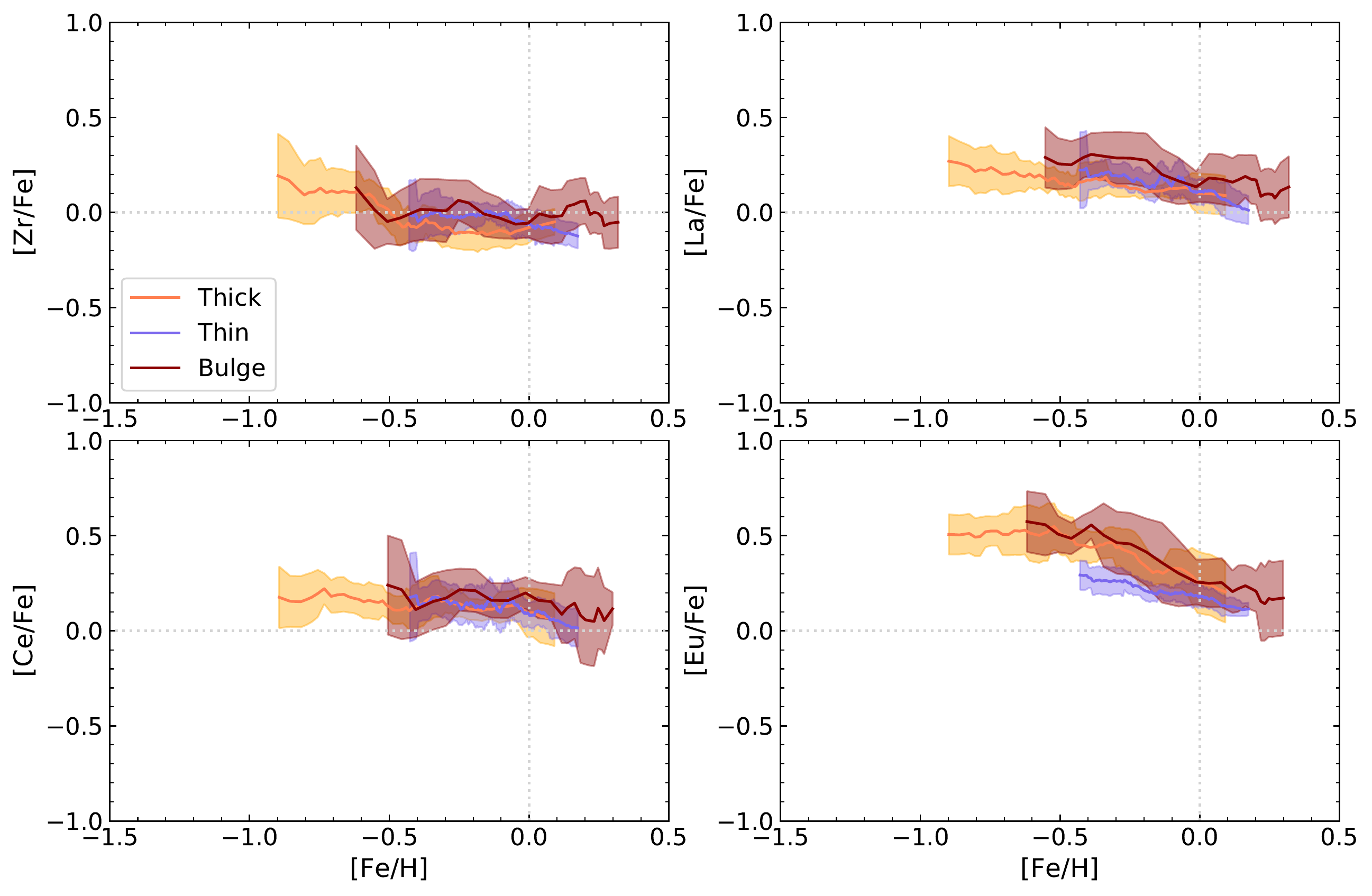}
 \caption{The running mean for the bulge (red solid line) compared with running mean of the thick disk (yellow solid line) and thin disk (blue solid line), with a 1$\sigma$ scatter (shaded regions, same colours as solid lines).}
              \label{fig: Running mean bulge + disk}%
\end{figure*}

In this section we compare our abundance-ratio trends, i.e. [X/Fe], for the bulge, the thin, and thick disks as a function of the metallicity for the s-process elements Zr, La, and Ce, and the r-process element Eu. In Fig. \ref{fig: Running mean bulge + disk} we directly compare the bulge population trends with those of the thin and the thick disk populations, determined in the same way in the present study. 

The bulge and the disks have very similarly shaped s-process element trends (Zr, La, Ce). The bulge trend of [La/Fe] have a slightly higher overall trend, especially at subsolar metallicities were [La/Fe] is $\sim 0.1$ dex higher than for the disk. We note that this is the opposite to findings in \citet{duong:19}. The metallicities of the bulge sample extends to slightly higher metallicities, pointing at a higher star formation rate of the bulge. Additionally, \citet{matteucci_ryde:19} shows that implementing a Salpeter like initial mass function (IMF), which favours massive stars compared to typical IMFs for the disk, better reproduce bulge abundances. 

%Tror vi skippar denna mening, det syns knappt tydligt ändå:
%Although, for [Zr/Fe] we see a possible increase of abundances around [Fe/H] $\sim$ 0.1, which could be the  AGB stars in the bulge.  

For [Eu/Fe], the thick disk is enhanced as compared with the thin disk, reminding us of an $\alpha$-element. The decreasing trend for metallicities larger than [Fe/H] $\gtrsim -0.4$ is a result of iron production by SN Ia after a time delay of roughly 100 Myr - 1 Gyr \citep{matteucci:1990,ballero:07}. The bulge traces the thick disk in the [Eu/Fe] abundance, pointing at the bulge having similar star formation rate as that of the thick disk. A plateau, or a knee, can be seen around metallicities of $\sim -0.4$ for both the thick disk and the bulge. A knee at higher metallicities than in the solar vicinity was already predicted for the bulge by \citet{matteucci:1990} and in general for systems with higher star formation rate than in the solar vicinity.

\begin{figure}
   \centering
\includegraphics[width=\hsize]{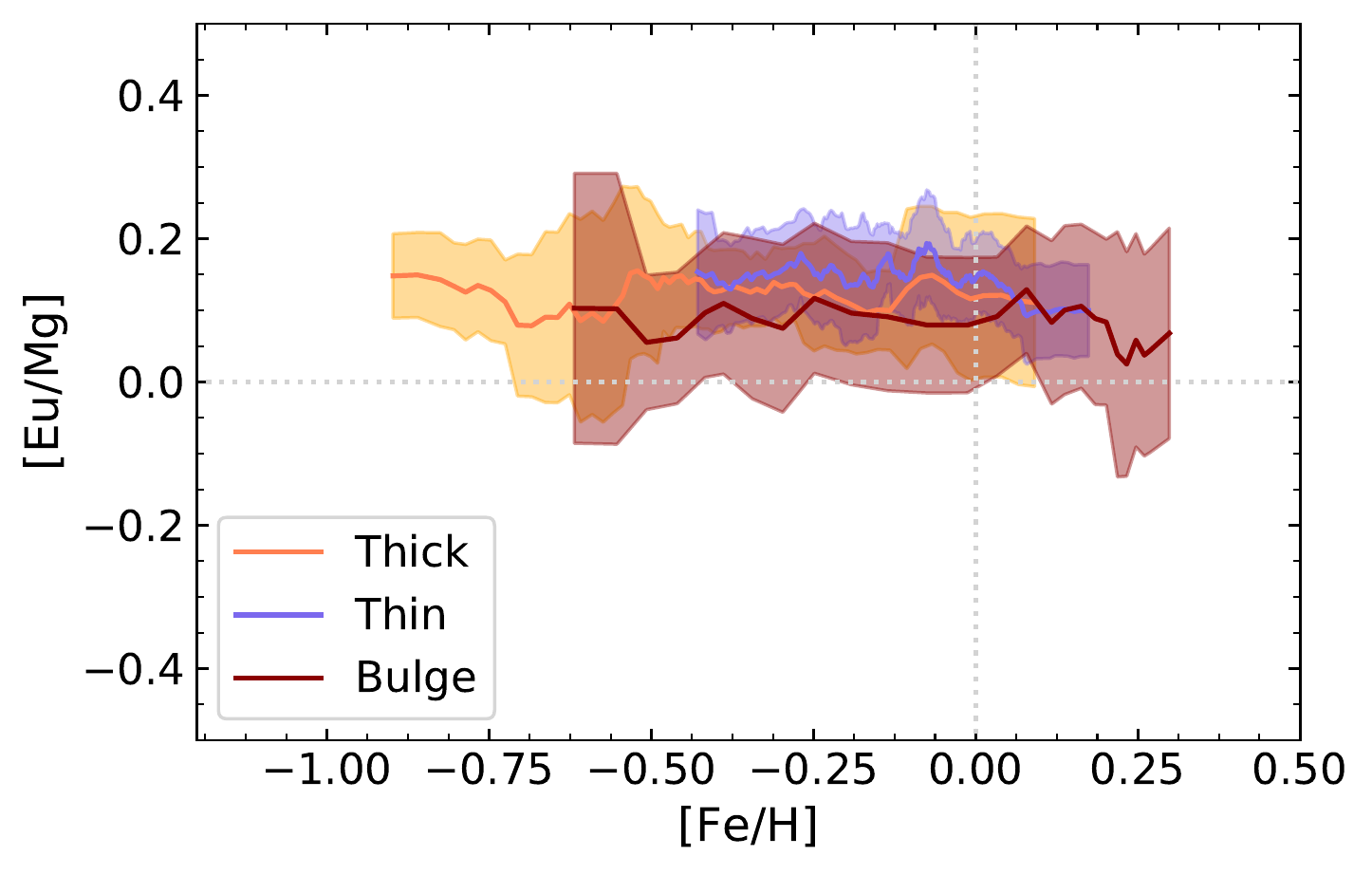}
 \caption{[Eu/Mg] abundances against [Fe/H] as running mean with a 1$\sigma$ scatter for the thin disk (blue), thick disk (yellow) and bulge (red).}
              \label{fig: [Eu/Mg]}%
\end{figure}
 
In Fig. \ref{fig: [Eu/Mg]} we compare Eu with the well-determined $\alpha$-element magnesium (Mg) \citep[from][]{Jonsson2017b}, by plotting [Eu/Mg], for the same stars. The resulting, mostly flat, trend of all populations is already expected from the [Eu/Fe] trend, pointing at Eu having a contribution from progenitors of similar timescales as that of progenitors producing Mg (i.e. SNe II). It has indeed been shown by \cite{travaglio1999} that SNe II progenitors with masses of 8-10 M$_{\odot}$ best reproduce the r-process enrichment in the Galaxy, and \citet{cescutti:06} showed that to reproduce the ratio of typical s-process elements, such as [Ba/Fe], at low metallicities, a r-process production of these elements in stars with masses ranging from 8 to 30 M$_{\odot}$ should be assumed. Nonetheless, the origin of r-elements is, as mentioned earlier, still debated \citep[see e.g.][]{sneden:2000,THIELEMANN2011346,cote:2018,siegel2019,kajino:2019}.

\begin{figure}
   \centering
\includegraphics[width=\hsize]{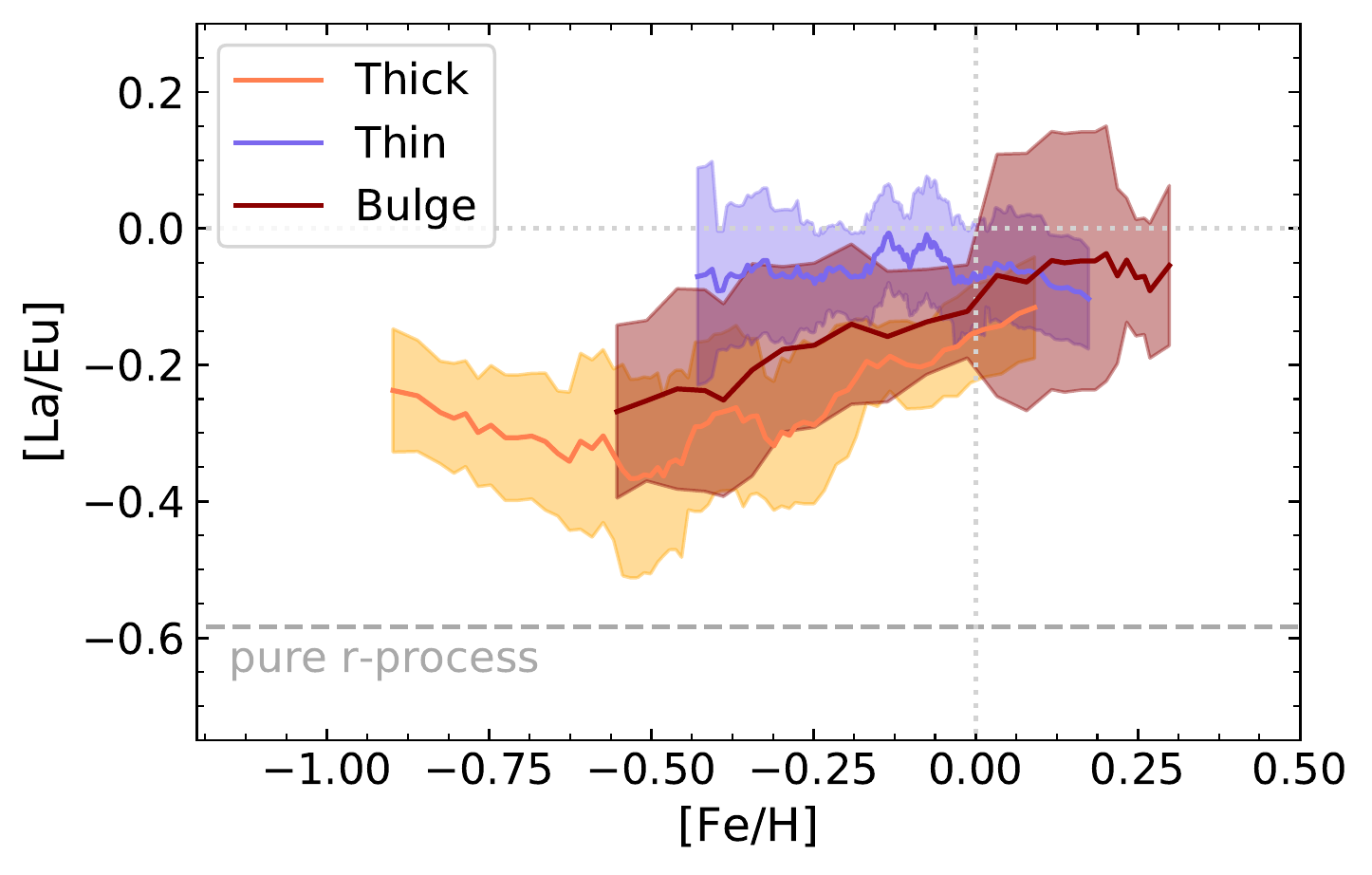}
 \caption{[La/Eu] abundances against [Fe/H] as running mean with a 1$\sigma$ scatter for the thin disk (blue), thick disk (yellow) and bulge (red). A pure r-process line is plotted, calculated using the values presented in \citet{bisterzo:2014}}
              \label{fig: [La/Eu]}%
\end{figure}

%[La/Eu] discussion
A way to disentangle the s- and r-process contribution throughout the evolution of the Galaxy is to compare an s-process dominated element with that of an r-process dominated one. We thus compare La, with an s-process contribution of 76~\%, to that of Eu with an r-process contribution of 94 \% \citep{bisterzo:2014}, plotted as [La/Eu] in Fig. \ref{fig: [La/Eu]}. A pure r-process line is added, using the values from \citep{bisterzo:2014}. The value of the pure r-process line is calculated by subtracting the predicted s-process abundance from the Solar System total values, i.e by treating the r-process as a residual \citep{bisterzo:2014}. 

The trends in Fig. \ref{fig: [La/Eu]} show that the r-process dominates more and more the production of neutron-capture elements when the metallicity decreases, reaching [La/Eu] $= -0.25$ for the bulge and [La/Eu] $= -0.4$ for the thick disk at [Fe/H] $\sim -0.5$. With regard to the large scatter at supersolar metallicities, we refrain any further interpretations of the bulge abundances at these metallicities. At around [Fe/H] $\sim -0.6$ the [La/Eu] thick disk trend levels off or even increases with lower metallicities. Whether this is significant or not is yet to be answered and observations of more stars in this metallicity range are needed. The in general higher [La/Eu] abundances of the bulge compared with those of the thick disk point at the bulge having either less r-process production (in turn, possibly a different amount of SNe II), or a higher s-process contribution (as seen previously in the [La/Fe]-trend) than that of the thick disk.

\subsection{First- and second-peak s-process elements}
\label{section: first second s process peak}

\begin{figure}
   \centering
\includegraphics[width=\hsize]{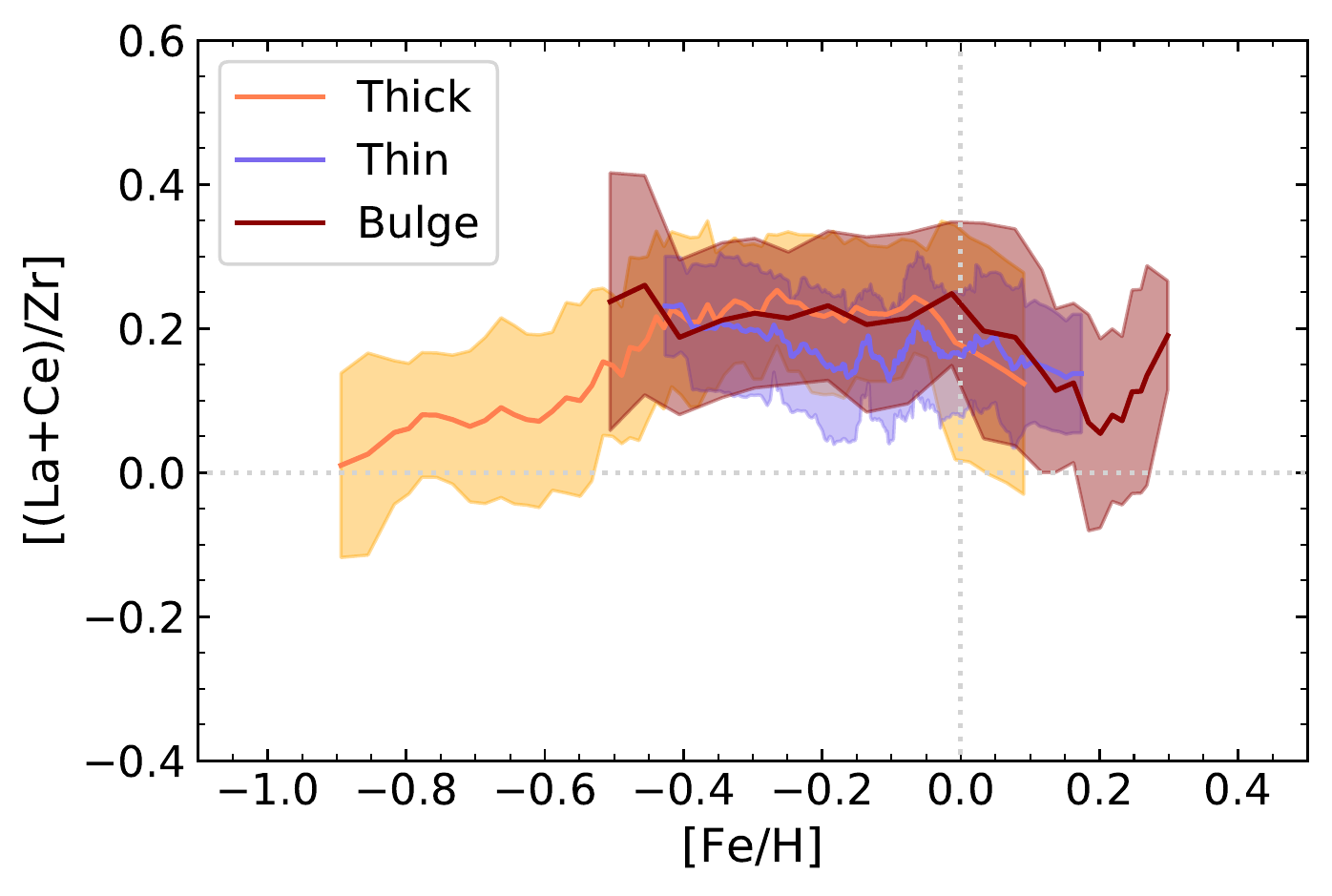}
 \caption{\label{fig: first second sprocess peak}  The abundances ratio of the second-peak s-process elements (La, Ce) and the first-peak s-process element (Zr) against [Fe/H] as running mean with a 1$\sigma$ scatter for the thin disk (blue), thick disk (yellow) and bulge (red).}
\end{figure}

In Fig. \ref{fig: first second sprocess peak}, the running mean of the ratio of the second-peak s-process elements (a mean of La and Ce) and the first-peak s-process element Zr are plotted against metallicity. The trend, elaborated on in the last paragraph of this section, can be explained by considering the stellar yields from \cite{2016ApJ...825...26K}, where low-mass AGB stars have a higher relative production of second-peak elements compared to the production of first-peak element.

The neutrons in the s-process come from two neutron sources, the ${}^{13}\text{C}(\alpha, n)^{16}$O- and the ${}^{22}\text{Ne}(\alpha, n)^{25}$Mg-reaction. The ${}^{13}$C source has a lower neutron density of roughly 10$^7$ neutrons/cm$^{3}$, whereas the neutron density for the ${}^{22}$Ne source is around 10$^{15}$ neutrons/cm$^{3}$. However, due to the longer timescales of the ${}^{13}$C reaction ($\sim 10^3$ years compared to $\sim 10$ years), the time integrated neutron flux for this neutron source is much higher than for the ${}^{22}$Ne source. Due to this, the ${}^{13}$C reaction builds up the heavier s-process elements, such as the second- (and third-)peak elements, whilst the ${}^{22}$Ne reaction is limited to producing the first-peak s-process elements \citep{karakaslattanzio:2014}.

Furthermore, \cite{2017ApJ...835...97B} elaborate on the importance of the size of the ${}^{13}$C-pocket in the s-process production. The ${}^{22}\text{Ne}$ reaction takes place only in initially more massive (AGB) stars of $>4$ M$_{\odot}$, due to the higher temperatures of these stars \citep{karakaslattanzio:2014}. This shrinks the ${}^{13}$C-pocket, resulting in a smaller quantity of s-elements to be expected, \textit{especially} the heavier ones. In short: heavier AGB stars produce relatively fewer second-peak elements compared to low-mass AGB stars, and the latter have a longer time delay. 

Another aspect to keep in mind is that at lower metallicities, the number of neutrons to the number of available ${}^{56}$Fe-seeds is higher, compared to higher metallicities, which enables the build-up of second-peak elements \citep{busso1999}.
%Consequently, at lower metallicities, the [second s peak/first s peak] abundance is higher, than at higher metallicities, something the could explain the increase in abundance towards lower metallicities [Fe/H] $<$ -0.5, for the thick disk (see Figure \ref{fig: first second sprocess peak}}.

In Fig. \ref{fig: first second sprocess peak} we first see an increasing trend in the thick disk for increasing metallicities, which turns over for solar metallicities and higher. Below solar metallicities (and above [Fe/H] $\sim -0.5$), all trends show an enrichment of second-peak as compared to first-peak elements.
%\textbf{For the metal-poor end of the thick disk ([Fe/H] $\lesssim -0.5$), we observe an increasing trend with increasing metallicity.}
This is therefore explained by the low-mass AGB stars which has not yet enriched the interstellar medium (ISM) at the time of the formation of the older thick disk stars, resulting in relatively low [(La+Ce)/Zr] abundances at early times. 

At solar metallicities, the disk populations does not show any clear differences. As for the bulge, it follows the trend of the thick disk more closely than that of the thin disk, at subsolar metallicities. At supersolar metallicities, the first-peak elements seem to increase in the bulge, possibly explained by a contribution of metal-rich AGB stars, producing a higher amount of first-peak elements \citep{2016ApJ...825...26K}.

%Results in \cite{travaglio2004} point at AGB stars not being the only important contributor to the s-element production in the thick disk, where the first peak elements could be produced in another process, such as Light Element Primary Production, or LEPP. Moreover, by including weak-s production from fast-rotating metal-poor stars, the abundance of the first peak elements can be increased \citep{2012A&A...538L...2F,2016MNRAS.456.1803F}. \textbf{ The paragraph on the LEPP sounds disconnected from the rest of the section and from the study. What is your point here? Same remark for the fast-rotators. Please reword to make a clear connection with the bulge/disk trends you present or drop this.}

%Kanske skippa detta?
% This seems to possibly be the case for the bulge and the thin disk.

\section{Conclusions}
In this work we have determined abundances of the neutron-capture elements Zr, La, Ce and Eu in 45 bulge giants and 291 local disk giants. The determination has been done using high-resolution spectra obtained with FLAMES/UVES (bulge sample) or either FIES or PolarBase (disk sample) and the analysis code SME. 

All spectra are evaluated over the wavelength region 5800 - 6800 Å and the careful, manual, definition of the continuum surrounding the spectral lines of interest in the spectra have been crucial in order to get high-precision abundances. Synthetic spectra allows the modelling and handling of blends in the spectra, as well as accounting for hyperfine splitting (in the cases of La, Eu). Isotopic shifts has also been taken into account by manually scaling the $\log(gf)$-values of the identified transitions in the line list (for the isotopes of Zr, Ce, Eu).

The stellar mass and metallicity are factors that contribute to, and affect, the s- and r-process production. Due to this, the enrichment of the ISM and the abundance of neutron-capture elements vary with time in the Galaxy, making them suitable probes for the Galactic chemical evolution. 

Our [Zr,La,Ce/Fe] bulge trends are in general flatter than those reported by previous studies, many of which are decreasing with higher metallicities. Such decreasing trends would suggest a higher r-process contribution to these elements in the bulge, while our flatter trends that have the same general shapes as our thick disk trends suggest more similar r/s-proportions in the creation of the neutron capture elements in the bulge and disks. The [La/Fe] bulge trend is $\sim 0.1$ dex higher compared with the disk, possibly indicating a higher s-process contribution in the bulge, compared with that of the disk.

For [Eu/Fe] we see a decreasing trend with increasing metallicities for both the disk and the bulge, with a plateau at around [Fe/H] $\sim -0.4$. This is very similar to the typical $\alpha$-element trend, and plotting [Eu/Mg] confirms this, pointing at the r-process having similar production rate as that of Mg (coming from SNe II).

For [La/Eu] we find that towards low metallicities, the abundances lay closer to the pure r-process line (reaching [La/Eu] $-0.4$ (disk) and $-0.25$ (bulge) at [Fe/H] $\sim -0.5$), indicating that the r-process was the dominating neutron-capture process at early times, both in the disk and the bulge. The results also point at either a) a different amount of massive stars or b) different contribution of the s-process, in the local thick disk and the bulge, where the [La/Eu] abundances seem to be systematically higher in the bulge than that of the thick disk. Since we compare abundances determined with the same method, for stars in the same evolutionary stage, the difference between the disk and the bulge in [La/Fe] could likely be real.

When plotting the ratio of the second- and first-peak s-process elements, [(La+Ce)/Zr], against metallicity we see that the bulge and the thick disk trends follow each other closely. We also show that, according to theoretical predictions by \citet{karakaslattanzio:2014}, low-mass AGB stars are needed to explain the enhancement of second-peak s-process abundances compared to first-peak s-process abundances.

To conclude, in general, our findings for Zr, Ce, Eu, point at the bulge having similar chemical evolution as that of the local thick disk, with similar star formation rate. On the other hand, our La trends for the bulge and the thick disk are offset by about 0.1: systematic effects could not be identified in our homogeneous analysis of the bulge and disk samples and further investigation is still required. Our results for the s-process elements differ substantially from previous works, where we find flatter trends. More bulge data would be needed to decrease the scatter and put further constraints on the bulge abundances. Additionally it would be useful to adopt the abundances to Galactic Chemical Evolution models to put further constraints on the evolution of the Galaxy and its components.

\begin{acknowledgements}
We would like to thank the referee, Mathieu Van der Swaelmen, for very insightful comments and suggestions that helped to improve this paper in many ways. 
This research has been partly supported by the Lars Hierta Memorial Foundation, and the Royal Physiographic Society in Lund through Stiftelsen Walter Gyllenbergs fond and M{\"a}rta och Erik Holmbergs donation.
H.J. acknowledges support from the Crafoord Foundation, Stiftelsen Olle Engkvist Byggm{\"a}stare, and Ruth och Nils-Erik Stenb{\"a}cks stiftelse. This work has made use of data from the European Space Agency (ESA) mission {\it Gaia} (\url{https://www.cosmos.esa.int/gaia}), processed by the {\it Gaia} Data Processing and Analysis Consortium (DPAC, \url{https://www.cosmos.esa.int/web/gaia/dpac/consortium}). Funding for the DPAC has been provided by national institutions, in particular the institutions participating in the {\it Gaia} Multilateral Agreement.
This publication made use of the SIMBAD database, operated at CDS, Strasbourg, France; NASA's Astrophysics Data System; and the VALD database, operated at Uppsala University, the Institute of Astronomy RAS in Moscow, and the University of Vienna.
\end{acknowledgements}

\bibliographystyle{aa}
\bibliography{mac_bib}

%%%%%%%%%%%%%%%%%%%%%%%%%%%%%%%%%%%%%%%%%%%%%%%%%%%%%%%%%%%%%%%%%%%%%%%%%%%%%%%%%%%%%%%%%
%%%%%%%%%%%%%%%%%%%%%%%%%% APPENDIX %%%%%%%%%%%%%%%%%%%%%%%%%%%%%%%%%%%%%%%%%%%%%%%%%%%%%
%%%%%%%%%%%%%%%%%%%%%%%%%%%%%%%%%%%%%%%%%%%%%%%%%%%%%%%%%%%%%%%%%%%%%%%%%%%%%%%%%%%%%%%%%
\begin{appendix}
\section{Additional tables}
% % ##################### Online material ##################
\begin{table}
\caption{Basic data for the observed bulge giants. The S/N per data point is measured by the IDL-routine \texttt{der\textunderscore snr.pro}, see \href{http://www.stecf.org/software/ASTROsoft/DER\textunderscore SNR}{http://www.stecf.org/software/ASTROsoft/DER\textunderscore SNR}.}
\begin{tabular}{l c c c c}
\hline
\hline
Star$^a$ & RA (J2000) & Dec (J2000) & $V$ & S/N\\
         & (h:m:s)    & (d:am:as)   &     & \\\hline
SW-09 & 17:59:04.533 & $-$29:10:36.53 & 16.153 & 16\\
SW-15 & 17:59:04.753 & $-$29:12:14.77 & 16.326 & 15\\
SW-17 & 17:59:08.138 & $-$29:11:20.10 & 16.388 & 11\\
SW-18 & 17:59:06.455 & $-$29:10:30.53 & 16.410 & 14\\
SW-27 & 17:59:04.457 & $-$29:10:20.67 & 16.484 & 13\\
SW-28 & 17:59:07.005 & $-$29:13:11.35 & 16.485 & 16\\
SW-33 & 17:59:03.331 & $-$29:10:25.60 & 16.549 & 14\\
SW-34 & 17:58:54.418 & $-$29:11:19.82 & 16.559 & 12\\
SW-43 & 17:59:04.059 & $-$29:13:30.26 & 16.606 & 16\\
SW-71 & 17:58:58.257 & $-$29:12:56.97 & 16.892 & 14\\
\hline
\end{tabular}
\label{tab:basicdata_bulge}
\tablefoot{This is only an excerpt of the table to show its form and content. The complete table is available in electronic form at the CDS.\\
\tablefootmark{a}{Using the same naming convention as \citet{lecureur:07} for the B3-BW-B6-BL-stars.}
}
\end{table}

\begin{table*}
\caption{Basic data for the observed solar neighbourhood giants. Coordinates and magnitudes are taken from the SIMBAD database, while the radial velocities are measured from the spectra. The S/N per data point is measured by the IDL-routine \texttt{der\textunderscore snr.pro}, see \href{http://www.stecf.org/software/ASTROsoft/DER\textunderscore SNR}{http://www.stecf.org/software/ASTROsoft/DER\textunderscore SNR}.}
\begin{tabular}{l l l l r r r l}
\hline
\hline
HIP/KIC/TYC & Alternative name & RA (J2000) & Dec (J2000) & $V$ & $v_{\mathrm{rad}}$ & S/N &  Source\\
        &                  & (h:m:s)    & (d:am:as)   &     & km/s\\
\hline
HIP1692 &           HD1690 & 00:21:13.32713 & $-$08:16:52.1625 &  9.18 &   18.37 &  114 & FIES-archive \\
HIP9884 &           alfAri & 02:07:10.40570 & +23:27:44.7032 &  2.01 &  $-$14.29 &   90 & PolarBase \\
HIP10085 &          HD13189 & 02:09:40.17260 & +32:18:59.1649 &  7.56 &   26.21 &  156 & FIES-archive \\
HIP12247 &            81Cet & 02:37:41.80105 & $-$03:23:46.2201 &  5.66 &    9.34 &  176 & FIES-archive \\
HIP28417 &          HD40460 & 06:00:06.03883 & +27:16:19.8614 &  6.62 &  100.64 &  121 & PolarBase \\
HIP33827 &           HR2581 & 07:01:21.41827 & +70:48:29.8674 &  5.69 &  $-$17.99 &   79 & PolarBase \\
HIP35759 &          HD57470 & 07:22:33.85798 & +29:49:27.6626 &  7.67 &  $-$30.19 &   85 & PolarBase \\
HIP37447 &           alfMon & 07:41:14.83257 & $-$09:33:04.0711 &  3.93 &   11.83 &   71 & Thygesen et al. (2012)\\
HIP37826 &           betGem & 07:45:18.94987 & +28:01:34.3160 &  1.14 &    3.83 &   90 & PolarBase \\
HIP43813 &           zetHya & 08:55:23.62614 & +05:56:44.0354 &  3.10 &   23.37 &  147 & PolarBase \\
\hline
\label{tab:basicdata_sn}
\end{tabular}
\tablefoot{This is only an excerpt of the table to show its form and content. The complete table is available in electronic form at the CDS.}
\end{table*}

\begin{table*}
\caption{Stellar parameters and determined abundances for observed bulge giants. [Fe/H] is listed in the scale of \citet{grevesse:2015}.}
\begin{tabular}{l c c c c c c c c}
\hline
\hline
Star & $T_{\mathrm{eff}}$ & $\log g$ & [Fe/H] & $v_{\mathrm{micro}}$ & A(Zr) & A(La) & A(Ce) & A(Eu)\\ 
\hline
SW-09 & 4095 & 1.79 & $-$0.15 & 1.32 & 2.79 & 1.09 & 1.72 & 0.75\\
SW-15 & 4741 & 1.96 & $-$0.98 & 1.62 &  ... &  ... & 1.51 &  ...\\
SW-17 & 4245 & 2.09 &    0.24 & 1.44 & 2.95 & 1.26 &  ... & 0.97\\
SW-18 & 4212 & 1.67 & $-$0.13 & 1.49 & 2.30 & 0.84 & 1.86 & 0.86\\
SW-27 & 4423 & 2.34 &    0.11 & 1.60 & 2.73 & 1.22 & 2.09 & 1.05\\
SW-28 & 4254 & 2.36 & $-$0.14 & 1.44 & 2.26 & 1.42 & 2.45 & 0.91\\
SW-33 & 4580 & 2.72 &    0.16 & 1.39 & 2.55 & 1.60 & 2.29 & 1.05\\
SW-34 & 4468 & 1.75 & $-$0.45 & 1.63 & 2.34 & 1.04 &  ... &  ...\\
SW-43 & 4892 & 2.34 & $-$0.77 & 1.84 &  ... & 0.75 &  ... & 0.28\\
SW-71 & 4344 & 2.66 &    0.39 & 1.31 & 3.10 & 1.77 &  ... &  ...\\
\hline
\label{tab:abundances_bulge}
\end{tabular}
\tablefoot{This is only an excerpt of the table to show its form and content. The complete table is available in electronic form at the CDS.}
\end{table*}

\begin{table*}
\caption{Stellar parameters and determined abundances for observed solar neighbourhood giants. [Fe/H] is listed in the scale of \citet{grevesse:2015}.}
\begin{tabular}{l c c c c c c c c}
\hline
\hline
HIP/KIC/TYC & $T_{\mathrm{eff}}$ & $\log g$ & [Fe/H] & $v_{\mathrm{micro}}$ & A(Zr) & A(La) & A(Ce) & A(Eu)\\ 
\hline
HIP1692  & 4216 & 1.79 & $-$0.26 & 1.55 & 2.20 & 0.97 & 1.41 & 0.51\\
HIP9884  & 4464 & 2.27 & $-$0.21 & 1.34 & 2.34 & 1.08 & 1.55 & 0.53\\
HIP10085 & 4062 & 1.44 & $-$0.32 & 1.63 & 2.32 & 1.03 & 1.48 & 0.51\\
HIP12247 & 4790 & 2.71 & $-$0.04 & 1.40 & 2.57 & 1.28 & 1.74 & 0.63\\
HIP28417 & 4746 & 2.56 & $-$0.25 & 1.40 & 2.24 & 1.02 & 1.39 & 0.52\\
HIP33827 & 4235 & 1.99 &    0.01 & 1.50 & 2.61 & 1.23 & 1.68 & 0.72\\
HIP35759 & 4606 & 2.47 & $-$0.15 & 1.42 & 2.23 & 1.06 & 1.54 & 0.74\\
HIP37447 & 4758 & 2.73 & $-$0.04 & 1.35 & 2.49 & 1.26 & 1.75 & 0.71\\
HIP37826 & 4835 & 2.93 &    0.07 & 1.24 & 2.68 & 1.33 & 1.79 & 0.73\\
HIP43813 & 4873 & 2.62 & $-$0.07 & 1.51 & 2.61 & 1.35 & 1.83 & 0.63\\
\hline
\label{tab:abundances_sn}
\end{tabular}
\tablefoot{This is only an excerpt of the table to show its form and content. The complete table is available in electronic form at the CDS.}
\end{table*}

%\section{Spectra}
%\label{appendix: disk spectra}
%\begin{figure}
%   \centering
%\includegraphics[width=\hsize]{bulge_lowSN.pdf}
% \caption{Same as Figure \ref{fig: spectrum bulge high S/N} but for the bulge star B3-B7 (S/N = 11)}
% \label{fig: spectrum bulge low S/N}%
%\end{figure}

%\begin{figure}
%   \centering
%\includegraphics[width=\hsize]{spectrum_arcturus.pdf}
% \caption{Same as Figure \ref{fig: spectrum bulge high S/N} but for the spectrum of the disk star Arcturus (HIP69673).}
% \label{fig: arcturus spectrum disk}%
%\end{figure}

%\section{Uncertainties}
%\label{appendix: uncertanties}
%\begin{figure}
%   \centering
%\includegraphics[width=\hsize]{uncertanties_distributions_disk.pdf}
% \caption{The abundance distribution from the uncertainty estimations using Arcturus ($\alpha$-Boo) as a typical star. A probability density function (orange) is fitted to the distributions (purple) and the standard deviation is noted in the upper left corner of every element. The measured abundance from the unaltered stellar parameters is marked with a dashed black line. The uncertainties here are estimated for a typical disk star.}
%              \label{fig: uncertanties disk dist}%
%\end{figure}

%\begin{figure}
%   \centering
%\includegraphics[width=\hsize]{uncertanties_distributions_bulge.pdf}
% \caption{Same as Figure \ref{fig: uncertanties disk dist} but with estimated uncertainties for a typical bulge star.}
%              \label{fig: uncertanties bulge dist}%
%\end{figure}

\end{appendix}

\end{document}